\documentclass[a4paper,11pt]{article}
\pdfoutput=1

\usepackage{jheppub}
\usepackage[T1]{fontenc}
\usepackage{amsmath,amsfonts,amssymb,textcomp}
\usepackage{slashed}
\usepackage{graphicx}
\usepackage[dvipsnames,table]{xcolor}
\usepackage{booktabs}

\def\beq{\begin{equation}}
\def\eeq{\end{equation}}
\def\beqa{\begin{eqnarray}}
\def\eeqa{\end{eqnarray}}

\def\ltap{\ \raise.3ex\hbox{$<$\kern-.75em\lower1ex\hbox{$\sim$}}\ }
\def\gtap{\ \raise.3ex\hbox{$>$\kern-.75em\lower1ex\hbox{$\sim$}}\ }
\def\ba{\begin{array}}
\def\ea{\end{array}}
\def\bea{\begin{eqnarray}}
\def\eea{\end{eqnarray}}
\def\bean{\begin{eqnarray*}}
\def\eean{\end{eqnarray*}}

\def\cL{{\cal L}}
\def\cA{{\cal A}}

\def\tr{{\rm Tr}}

\newcommand{\UY}{U(1)_Y}
\newcommand{\UBL}{U(1)_{B-L}}

\newcommand{\Upaxt}{U(1)_\text{ax}^{\prime \, c_V^t }}
\newcommand{\Uem}{U(1)_\text{EM}}
\newcommand{\UBpL}{U(1)^\prime_{B+L}}
\newcommand{\UL}{U(1)^\prime_{L}}
\newcommand{\gZp}{g_{Z'}}
\newcommand{\aem}{\alpha_\text{em}}
\newcommand{\mZp}{m_{Z'}}
\newcommand{\GZp}{\Gamma_{Z'}}
\newcommand{\mX}{m_\chi}
\newcommand{\mf}{m_f}
\newcommand{\gX}{g_\chi}

\newcommand{\cAf}{c_A^f}
\newcommand{\sSD}{\sigma_p^\text{SD}}
\DeclareMathOperator{\Tr}{Tr}

\subheader{\footnotesize
\vspace*{-3.7em}
\begin{flushright}
CERN-TH-2017-142\\
PITT-PACC-1705
\end{flushright}
}

\title{On Dark Matter Interactions with the Standard Model through an Anomalous $Z'$}

\author[a]{Ahmed Ismail,}
\author[b,c]{Andrey Katz}
\author[c]{and Davide Racco}

\affiliation[a]{Pittsburgh Particle Physics, Astrophysics, and Cosmology Center,\\Department of Physics and Astronomy, University of Pittsburgh, 3941 O'Hara St.,\\Pittsburgh, PA 15260, USA}
\affiliation[b]{Theory Division, CERN, CH-1211 Geneva 23, Switzerland}
\affiliation[c]{D\'{e}partement de Physique Th\'{e}orique and Center for Astroparticle Physics (CAP),\\Universit\'{e} de Gen\`{e}ve, 24 quai Ansermet, CH-1211 Gen\`{e}ve 4, Switzerland}

\emailAdd{aismail@pitt.edu}
\emailAdd{andrey.katz@cern.ch}
\emailAdd{davide.racco@unige.ch}

\abstract{We study electroweak scale Dark Matter (DM) whose interactions with baryonic matter are mediated by a heavy 
anomalous $Z'$.
We emphasize that when the DM is a Majorana particle, its low-velocity annihilations are dominated by loop suppressed
annihilations into the gauge bosons, rather than by p-wave or chirally suppressed annihilations 
into the SM fermions. Because the 
$Z'$ is anomalous, these kinds of DM models can be realized only as effective field theories (EFTs) with a well-defined 
cutoff,  where heavy spectator fermions restore gauge invariance at high energies. We formulate 
these EFTs, estimate their cutoff and properly take into account the effect of the Chern-Simons terms one 
obtains after the 
spectator fermions are integrated out.
We find that, while for light DM collider and direct detection experiments usually
provide the strongest bounds, the bounds at higher masses are heavily dominated by indirect detection experiments, 
due to strong annihilation into $W^+W^-$, $ZZ$, $Z\gamma$ and possibly into $gg$ and $\gamma\gamma$.  
We emphasize that these annihilation channels are generically significant because of the
structure of the EFT, and therefore these models are prone to strong indirect detection constraints.
Even though we focus on selected $Z'$ models for illustrative purposes, our setup is completely generic
and can be used for analyzing the predictions of any anomalous $Z'$-mediated DM model with arbitrary charges.
}

\begin{document} 
\maketitle
\flushbottom

\section{Introduction}
\label{sec:intro}

While experimental evidence for dark matter (DM) has been well established for decades, the
precise nature of DM remains unknown to this day. Searches for non-gravitational interactions
of DM use a broad array of techniques to test different models, from tabletop experiments
probing axions to ton-scale detectors that are sensitive to DM much heavier than the proton.
With many proposed mechanisms for realizing DM in nature, the development of new DM detection
methods is a highly active field. Simultaneously, theoretical advances have guided experiment
by exploring frameworks that incorporate DM naturally into extensions of the Standard Model (SM).

From a cosmological perspective, thermal freeze-out is one of the simplest ways to account for
the observed abundance of DM. The idea that DM is a thermal relic is also attractive
phenomenologically, as it implies a connection between the DM relic density and the strength
of potential signatures. Now, it is well known that thermal relic DM candidates are 
subject to strong model-independent constraints, based on unitarity considerations of DM 
annihilation~\cite{Griest:1989wd}. In particular, a thermal relic DM particle cannot have 
a mass exceeding $\sim 300$~TeV. In practice, this limit is not easy to saturate and within concrete 
models the mass bound on a thermal relic is expected to be significantly more modest. 
While the DM is not necessarily a thermal relic and a plethora of other consistent candidates 
have been studied in the literature (for reviews 
see~\cite{Bertone:2004pz,Petraki:2013wwa,Baer:2014eja}), 
thermal relics are still appealing candidates, both for theoretical reasons and because 
these bounds tightly constrain their allowed parameter space and in principle allow 
a thorough study with collider, direct and indirect detection experiments.

Within the thermal freeze-out scenario, it is well known that a particle that interacts weakly and is
near the electroweak scale $\sim 1$~TeV would provide approximately the observed DM relic abundance.
These so-called weakly interacting massive particles (WIMPs) are still the most popular DM 
thermal relic candidates, in spite of the fact that large parts of their parameter space 
have already been ruled out both by direct and indirect detection experiments. Specifically, while
the XENON1T experiment provides the strongest direct detection bounds on 
WIMPs~\cite{Aprile:2017iyp}, indirect limits arise 
from the observations of Dwarf Spheroidal Galaxies (dSph) by 
Fermi-LAT~\cite{Drlica-Wagner:2015xua,Ackermann:2015zua} 
and of diffuse 
$\gamma$ rays from the Galactic Center by HESS~\cite{Abdallah:2016ygi}. 

In this regard, there is good motivation 
to consider a broader set of thermal relic candidates, beyond the ``standard'' WIMP paradigm. 
This led to the emergence of dark matter effective field theories 
(EFTs)~\cite{Goodman:2010yf,Bai:2010hh,Goodman:2010ku}, 
which posit higher-dimension
operators between DM and SM states. While dark matter EFTs, especially in their non-relativistic
formulations~\cite{Fan:2010gt,Fitzpatrick:2012ix}, are relevant for interpreting the results of direct detection
experiments, they are unable to accurately describe the physics of processes where the momentum transfer
is comparable to the EFT cut-off scale, as is common in collider searches~\cite{Busoni:2013lha,Buchmueller:2013dya,Busoni:2014sya,Busoni:2014haa}.
The requirement to consider ultraviolet completions of dark matter EFTs in these regimes subsequently
resulted in the development of simplified DM models. A typical simplified DM model extends the SM by
a DM candidate as well as a mediator that communicates between the SM and dark sectors.

Even though it is hard to believe that any simplified model accurately describes 
all physics beyond the SM, the essential idea is that the key ingredients that
determine the experimental signatures related to the DM should be captured correctly by these 
models. For these purposes simplified models must 
be able to make proper predictions for the thermal relic abundance, 
direct detection experiments, neutrino telescopes, $\gamma$ ray telescopes and 
collider experiments, such as the LHC or a future 100~TeV machine. We will closely
investigate this requirement in our work in the context of simplified models with spin-1 
mediators.

The idea that the interaction between the SM particles and DM is mediated by a heavy 
neutral spin-1 boson, that we will further call $Z'$, is not new. 
Refs~\cite{An:2012va,Frandsen:2012rk,Arcadi:2013qia,Chu:2013jja,Dudas:2013sia,Fairbairn:2014aqa,Lebedev:2014bba,Alves:2015pea,Alves:2015mua,Kahlhoefer:2015bea} 
form merely 
a partial list of the related contributions. In this particular work we will concentrate on
a Majorana fermion DM candidate whose interactions with the SM are mediated by the heavy $Z'$, 
corresponding to a symmetry that appears to be {\it anomalous} at the electroweak scale. 
Anomaly cancellation at  high scales is necessary for the overall consistency of the theory, 
as well as for more practical purposes, most notably the calculation of the couplings 
of the $Z'$ to the SM gauge bosons, which in turn largely determine the DM signatures in 
indirect detection experiments. This point has been recently emphasized in~\cite{Jacques:2016dqz}.
Moreover, many of the $Z'$ models employed in describing the results of LHC searches, including 
the ``axial'' $Z'$ model, are anomalous~\cite{Boveia:2016mrp,Albert:2017onk}. 
All such ``anomalous'' theories must descend from the UV complete ones, 
where the anomalies are either canceled
by spectator fermions~\cite{Ismail:2016tod}, or via the Green-Schwarz 
mechanism~\cite{Green:1984sg,Coriano:2005own,Armillis:2008vp}. As 
has been recently shown in Ref.~\cite{Ellis:2017tkh}, these spectator fermions can be 
potentially responsible for non-trivial collider signatures and can be more easily 
accessible at the LHC
than the DM itself.    

In this paper we will take a different approach. In fact, it is not always 
necessary to analyze a 
{\it full UV-complete model} to make important predictions for DM signatures in relevant 
experiments. In particular, we will be especially interested in the anomalous $Z'$ couplings 
to the SM gauge bosons. These couplings determine the annihilation cross sections of DM 
into SM gauge bosons, affecting the $\gamma$ ray fluxes from dSph 
and the Galactic Center, as well as signals in neutrino telescopes. To calculate these 
observables, it is sufficient to consider an EFT with the anomalous 
$Z'$ after the heavy spectators have been integrated out. 

In fact, EFTs with low-energy anomalies from integrating out heavy chiral fermions have been 
considered as early as the 1980s, mostly in the context of 
the SM without the top quark~\cite{DHoker:1984mif,DHoker:1984izu}. Indeed the 
$SU(2)\times U(1)$ electroweak symmetry is anomalous in the absence of the top quark and 
should be analyzed as an effective field theory with extra degrees of freedom with couplings 
which 
compensate for the loss of gauge invariance at the 1-loop level.  This approach was further 
generalized by Preskill in Ref~\cite{Preskill:1990fr}.
More recently, the influence of anomalous $Z'$ couplings to the SM gauge bosons has been 
studied in the context of 
DM~\cite{Mambrini:2009ad,Dudas:2009uq,Dudas:2013sia,Domingo:2013tna,Arcadi:2017jqd}. 

In this work we essentially take the same approach. We formulate ``simplified models''
of DM with anomalous $Z'$ mediators as consistent effective field theories with a cutoff $\Lambda$. 
We will show, in agreement with the results of~\cite{Preskill:1990fr}, that this cutoff can 
be much heavier than the mass of the $Z'$ and therefore the non-decoupling effects of the 
heavy spectators can be efficiently captured by the EFT, without explicitly considering 
these fermionic degrees of freedom. As expected, this EFT uniquely determines the couplings 
between the heavy $Z'$ and the SM gauge bosons in which we will be interested~\cite{Chu:1996fr,Bilal:2008qx}. 

Because we are considering an EFT, we will find that some of 
our amplitudes, including $\chi \chi \to V V$, where $V$ is a SM gauge boson,
grow {\it quadratically} with energy. This should not be surprising, as the EFT necessarily contains higher dimensional operators, without which 
gauge invariance is lost. The growth of such amplitudes is tamed at the scale 
$\Lambda$, where the spectator fermions appear. 

In this work we explicitly calculate the annihilation rates of the dark matter into the SM 
gauge bosons and estimate the bounds, associated with these rates. We choose as examples several
anomalous $Z'$ models, that illustrate some generic patterns. We emphasize, that while the 
concrete bounds are always model dependent, the techniques that we illustrate here are completely
generic and can be used in any EFT with an anomalous $Z'$ mediator.
  
We find that for DM heavier than $\sim 200$~GeV, these higher dimensional operators dictate that DM annihilation at low velocities is dominated by final states involving gauge bosons. This results in considerable bounds from indirect detection experiments. At larger velocities, such as at DM freeze-out, $p$-wave annihilation into fermions overcomes these operators, which are loop suppressed, and so the DM relic abundance calculation is mostly 
unaffected by the requirement of anomaly cancellation. We also compare our new indirect detection constraints with direct detection and collider limits.
We find that for heavy DM, 
$\gamma$ ray and neutrino telescopes (depending on the concrete model) 
provide the strongest bounds on anomalous $Z'$ simplified DM models.

The remainder of this paper is organized as follows. In Sec.~\ref{sec:eftanomaly}, 
we describe the effect of integrating out heavy fermions in a consistent theory, 
yielding an EFT with apparent anomalies at low energy scales. We focus on the induced loop level operators corresponding to these anomalies, and show that the maximum EFT cutoff can be significantly larger than the $Z'$ mass. Then in Sec.~\ref{sec:models} we specialize to the case of simplified models of DM and motivate a selection of toy models which serve to illustrate the effects of the higher dimensional operators on physical observables. Sec.~\ref{sec:dm} contains the experimental bounds on these simplified models, paying particular attention to the impact of loop-induced DM annihilation to gauge bosons on indirect detection constraints. We briefly discuss some 
limitations  of our analysis in Sec.~\ref{sec:validity}, 
including the assumption that the spectator fermions are heavy. 
Sec.~\ref{sec:conc} contains our conclusions. Some important numerical results of our 
calculation are relegated to the Appendix.

\section{Low-Energy Effective Theory}
\label{sec:eftanomaly}

In this section we will review building of an EFT for a new gauge group
which appears to be anomalous at low energies. We will introduce  
the new couplings that should 
necessarily be introduced to restore gauge invariance of the full theory.
We will also describe how the couplings between the exotic and 
SM gauge bosons should be calculated, from anomaly considerations.
Throughout we will closely follow 
the original work by Preskill~\cite{Preskill:1990fr} (as well as slightly more 
detailed handwritten notes~\cite{Preskill:1983}).
We also borrow some results
from more recent works~\cite{Anastasopoulos:2008jt,Racioppi:2009yxa} 
that made practical use of these results 
in a slightly different context
of MSSM augmented with anomalous $Z'$s. 
The reader familiar with this subject may safely skip 
this section and proceed directly to Sec.~\ref{sec:models}, where we explain in detail the DM
models that we consider, and Sec.~\ref{sec:dm}, where we present our results.  

The EFTs that we are describing here can be thought of as descending from 
fully gauge invariant theories with spontaneously
broken gauge symmetry, after some heavy fermions have been integrated out. 
Given that these fermions are chiral and get masses from gauge 
symmetry breaking, like the fermions in the SM, 
the theory below the scale of the heavy fermions appears to be anomalous. 
In fact, this is exactly what happens in the SM if we integrate out the top quark, 
as it is the heaviest fermion of the SM~\cite{D'Hoker:1984ph,D'Hoker:1984pc}. 
Although the full SM is perfectly 
anomaly free, as one would expect from a consistent gauge invariant theory, 
integrating out the top leaves both the hypercharge and the $SU(2)_L$ symmetries 
anomalous, as well as giving rise to the $SU(2)_L \times U(1)_Y$ mixed anomaly. 

The basic procedure of canceling the anomalies in this low energy EFT 
comes at the price of introducing
non-renormalizability. To see this in a working example, let us consider first 
the $U(1)'^3$ triangle in an anomalous  Abelian theory.
Since the theory is anomalous, gauge invariance is lost and under the 
$U(1)'$ gauge transformation $A_\mu \to A_\mu + \partial_\mu \omega / g'$ the effective action 
transforms as
\beq\label{eq:U1trans}
\delta_\omega \Gamma = \frac{g'^2}{ 96 \pi^2} \sum_i Q_i^3 \int d^4 x\, \omega\, 
\epsilon_{\mu \nu \rho \sigma} F^{\mu \nu} F^{\rho \sigma}~.
\eeq  
where $F$ is the field strength associated with $A$. Here $g'$ stands for the ``gauge'' coupling of the $U(1)'$ and the sum runs 
over all the fermions that are charged under the $U(1)'$. 
This transformation simply manifests the fact that in an anomalous theory the 
gauge invariance is lost. There will also be mixed anomalies between the $U(1)'$ and the SM gauge groups, which we consider below.

In our example, the $U(1)'$ gauge invariance 
can be easily restored by introducing a scalar $a$ that transforms under a 
gauge transformation as $a \to a + v\, \omega $, where $v$ stands for the 
scale of the $U(1)'$ breaking, or, equivalently $v \equiv  m_{Z'}/g'$. Then,
the transformation~\eqref{eq:U1trans} can be restored by introducing the
following term:
\beq\label{eq:U1counter}
\cL = - \frac{g'^2}{96\pi^2} \sum_i Q_i^3 
\epsilon_{\mu\nu\rho\sigma} 
\frac{a F^{\mu\nu} F^{\rho \sigma}}{v}~.
\eeq

Even though~\eqref{eq:U1counter} appears to cancel the anomaly 
with a new degree of freedom, 
this term is merely a Wess-Zumino counterterm that we have added to the action 
and $a$ is not a genuine degree of freedom. First, it is worth noticing that 
in spite of the form of Lagrangian term~\eqref{eq:U1counter}, the total 
Lagrangian is \emph{independent} of the field $a$ and depends only on its 
derivative
$\partial_\mu a$. This fact becomes manifest if we perform a 
rotation on the fermions
$\psi_i \to e^{-i Q_i a/v} \psi_i$.\footnote{To avoid confusion we will assume 
two-component fermions in this notation. One can always use them to construct the four-component 
 fermions with an appropriate use of the projection operators.}
While such a rotation eliminates the 
term~\eqref{eq:U1counter}, the path integral measure transforms non-trivially  
under this rotation, inducing a term in the effective action that looks like 
$\sim \partial_\mu a \psi^\dagger \bar \sigma^\mu \psi$.

\begin{figure}
\centering 
\includegraphics[width = .99\textwidth]{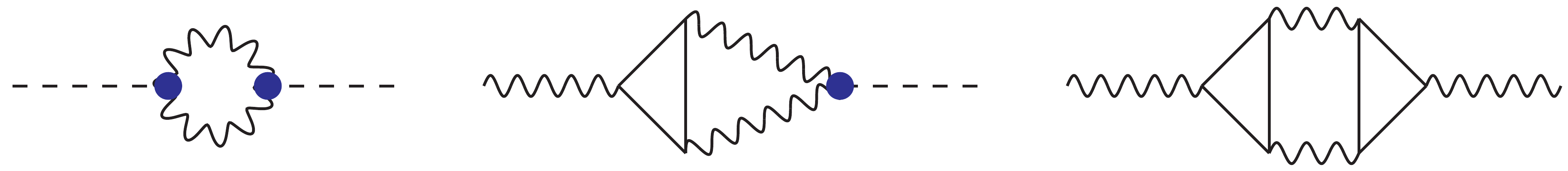}
\caption{Kinetic term for the field $a$ and the mass term for the $Z'$
gauge boson induced by the radiative corrections in the $U(1)$ anomalous theory. Note that each 
blob corresponds to the couplings~\eqref{eq:U1counter} and is therefore naturally of the size 
of the fermionic triangle loop. }
\label{fig:radcor}
\end{figure}

The kinetic term of the field $a$, which should also be gauge invariant, is of the form
\beq\label{eq:kineticStuck}
\cL = \frac{1}{2} \left( \partial_\mu a - g' v Z'_\mu \right)^2
\eeq
Even if we start from a theory that does not have this term, it is induced 
radiatively by the diagrams depicted in Fig.~\ref{fig:radcor}, 
similarly to the Green-Schwarz mechanism.  

This Lagrangian is nothing but a $U(1)$ theory that 
has been higgsed via a St\"uckelberg mechanism. 
In the unitary gauge the scalar degree of freedom $a$ can be set (locally) 
to zero at any point in space, leaving us simply with an effective  
theory of the massive gauge bosons with anomalous fermionic field content. 

Of course, our effective theory cannot be extrapolated to infinitely 
high energies, and the calculability requirement sets the cutoff of the theory. 
In any non-unitary gauge, the presence of the cutoff is evident from the 
term~\eqref{eq:U1counter} in the Lagrangian, while in the unitary 
gauge we can see it from the bad UV behavior of the two-point function of the 
$Z'$. 
In order to estimate the cutoff of the effective theory, we should remember that loop effects, similar to those that produce the term~\eqref{eq:kineticStuck}
(see Fig.~\ref{fig:radcor}), will also produce  terms that look like 
\beq\label{eq:PertExpan}
\sim \frac{1}{(4\pi v)^{p-2}} \frac{1}{v^{p-2}} \left( \partial_\mu a - g' v Z'_\mu 
\right)^p
\eeq
for every power $p \geq 2$. In order to have a consistent EFT, each order in the 
perturbative expansion~\eqref{eq:PertExpan} should be smaller than its 
predecessor such that the expansion is 
valid.\footnote{While we will hereafter dub this expansion as a ``loop expansion'', 
it is important to keep in mind that the couplings from Eq.~\eqref{eq:U1counter} 
and the fermion
loop are of the same order of magnitude, similarly to the Green-Schwarz mechanism.}
Taking this requirement into account 
(see~\cite{Preskill:1990fr} for the details of this derivation) one finds the following 
cutoff estimation 
of the EFT: 
\beq\label{eq:cutoff}
\Lambda \sim \frac{64 \pi^3 m_Z'}{\left| g' \sum_i Q_i^3\right|}~.
\eeq    

Now we can extend this logic to models with more complicated gauge 
symmetries and mixed anomalies between the $U(1)'$ and the non-Abelian 
gauge groups. This is exactly the situation in which we are interested,
where the anomalous $Z'$ couples to the DM, and the mixed anomaly will 
eventually determine the strength of its interaction with the SM gauge 
bosons. 

The treatment of the mixed anomalies will follow a similar logic 
to one we used in the $U(1)'$ case. 
Let us consider $U(1)' \times SU(N)$ 
symmetry with a mixed anomaly 
\beq
\sum_i \tr (t^a t^b Q_i) = \cA \, \delta^{ab}
\eeq
where $t^a$ are the generators of $SU(N)$ 
The matrix element of this theory between the $Z'$ and the $SU(N)$ gauge bosons
is nominally divergent, signaling that the theory is non-renormalizable, because
there is no tree level coupling between the $Z'$ and the $SU(N)$ gauge bosons. 

In this case the form of the anomalous transformation is slightly less straightforward
to derive. However, it can be obtained by invoking the Wess-Zumino consistency 
condition~\cite{Wess:1971yu}. 
Under $U(1)'$ and $SU(N)$
transformations with transformation parameters $\omega_1$ and $\omega_N$, 
respectively, the action transforms as:
\beqa\label{eq:trans1}
\delta_{\omega_1} \Gamma  & = & C_1 \frac{g_N^2}{16\pi^2} \cA \Tr \int d^4 x \omega_1
\left( F_N^{\mu \nu} \tilde F_{N\mu \nu}\right)\\ \label{eq:trans2}
\delta_{\omega_N} \Gamma & = &  C_N \frac{g'_1 g_N}{ 8 \pi^2} \cA \int d^4 x \tilde F_1^{\mu \nu}
\Tr \left( \omega_N \partial_\mu A_{N \nu}\right)
\eeqa
Note that we have only kept the components of the transformations that correspond to
the mixed anomaly, and their sum is fixed by the Wess-Zumino consistency condition to be 
$C_1 + C_N =1$. In particular, the presence of the mixed anomaly means that we cannot 
simultaneously
have $SU(N) \times U(1)'$ gauge invariance, since either $C_1$ or $C_N$ must be non-zero.
Conversely, the orthogonal combination $C_1 - C_N$ is 
unconstrained. This combination
depends on the Wess-Zumino counterterm which shifts the value of 
$C_1 - C_N$. That counterterm can be added to the action with an arbitrary 
strength.

In an arbitrary $SU(N) \times U(1)'$ gauge theory, there is no {\it a priori} motivation
to choose particular values of $C_1$ and $C_N$, which can be used, in particular, to insist
that the anomaly preserves either $U(1)'$ or $SU(N)$ gauge invariance.
However, in the SM augmented with the anomalous 
$Z'$ the situation is different. While we expect that at the scale $\Lambda$, or below, 
the spectator fermions restore the full gauge invariance, we should also insist that even 
below the spectator fermion scale the SM electroweak gauge group is exactly gauge invariant.
Otherwise, the anomaly would affect electroweak gauge group.
This requirement will set for us the coefficient in front of the Wess-Zumino 
counterterm and consequently the value of the combination $C_1 - C_N$. Indeed, using the freedom 
to set $C_1-C_N$ we can always choose the counterterm such that 
either $C_1 = 0$, namely  require the $U(1)$ gauge invariance, or 
$C_N = 0$, which would mean that the $SU(N)$ is gauge invariant. 
In the SM with the $Z'$, we will have to insist that the corresponding gauge transformations 
of the $SU(2)_L \times U(1)_Y$ vanish, but {\emph not} of the $U(1)'$.

This requirement of the gauge transformation of the electroweak group is crucial for our further 
calculation. It further removes any ambiguities in the calculation of the $Z'$ vertex with any pair
of the SM gauge bosons. We will now outline this calculation. For illustrative purposes we will 
assume here an unbroken electroweak symmetry with massless fermions. Of course in the SM the 
electroweak symmetry is broken and we consider the effects of the breaking, including fermion masses 
and the contributions from the Nambu-Goldstone bosons of $SU(2)_L\times U(1)_Y$,
in our explicit calculation in Sec.~\ref{sec:dm}. However, eventually the spontaneous electroweak symmetry breaking is a minor effect that does not change the picture conceptually. 

\begin{figure}[t]
\centering
\includegraphics[width = 0.99\textwidth]{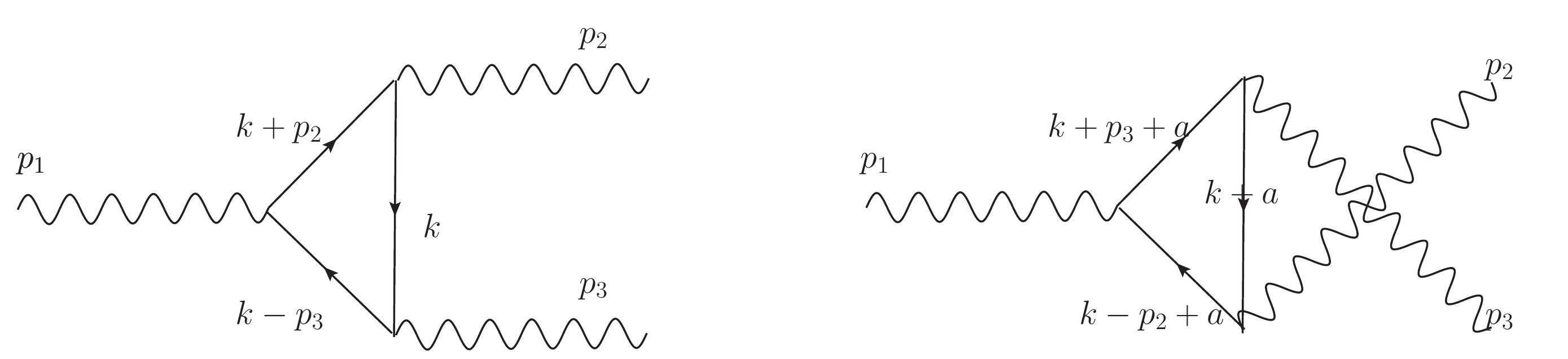}
\caption{Diagrams relevant for the $Z'VV$ vertex function calculation. The integration 
variable $k$ of one diagram can be shifted with respect to another by an arbitrary momentum 
$a$. }
\label{fig:VertexFunction}
\end{figure}

The calculation of the $Z'VV$ vertex function involves the calculation of the pair 
of diagrams shown in Fig.~\ref{fig:VertexFunction}, 
where we understand the sum over all the fermions  charged under the $U(1)'$ and the relevant SM gauge group. In an anomaly free theory one can 
always shift the integration momentum of one diagram with respect to the other by an 
arbitrary momentum $a_\mu$, 
without changing the finite answer. This is no longer true in an anomalous 
theory. As we will immediately see the momentum shift $a$ is \emph{not arbitrary} in 
our setup, and in fact for a given Wess-Zumino counterterm it will be completely determined 
by the required gauge invariance of the EW group. 

Our objective is to make sure that 
only the gauge transformation of the effective action with 
respect to the $U(1)'$ does not vanish, which is equivalent to the requirement that the Ward 
(Slavnov-Taylor) identities for the EW gauge group hold. Namely, in the case of the unbroken 
$SU(2)_L \times U(1)_Y$ we get 
\beq\label{eq:SimpleWard}
p_\mu \Gamma_\text{3-point}^{\mu \nu \rho} = 0
\eeq
when $p_{\mu}$ is the momentum of the SM gauge boson, which would correspond to $p_2$ and 
$p_3$ in Fig.~\ref{fig:VertexFunction}. When we will move to the broken EW symmetry,  
Eq.~\eqref{eq:SimpleWard} will be accompanied by an extra piece, corresponding to the 
Goldstone boson contribution, that we will take care of later. 

When we are dealing with the mixed anomaly, the expression that one 
gets in~\eqref{eq:SimpleWard} is $a_\mu$-dependent. Because the anomalies do not 
cancel out, each separate term of  $\Gamma^{\mu \nu \rho}_\text{3-point}$ is nominally 
linearly divergent. Therefore, because of the freedom to shift the integration momentum 
by $a_\mu$, we expect the Ward identity to have a form 
\beq\label{eq:FullWard}
p_\mu \Gamma^{\mu \nu \rho} \sim \int d^4 k \left( f^{\mu \nu \rho}(k_\sigma + a'_\sigma) 
- f^{\mu \nu \rho}(k_\sigma) \right)~,
\eeq 
with the leading term of $f^{\mu\nu\rho}$ in $k$ leading to the linear divergence of the 
integral. 
Note, that generically the shift momentum $a'$ need not be exactly equal to $a$, but rather  
can involve a linear combination of the external momenta as well.
After expanding the first term we find that the result does not vanish, but rather 
reduces to a surface 
term
\beq
2 i \pi^2 a'^\sigma \lim_{k^2 \to \infty} k^2 k_\sigma f^{\mu \nu \rho} (k)~, 
\eeq
which is finite and $a$-dependent. There is no choice of $a_\mu$ to set the Ward identities
in Eq.~\ref{eq:FullWard} to zero simultaneously for all three incoming momenta $p_i$ in
Fig.~\ref{fig:VertexFunction}. 
However, there is \emph {always a choice of $a_\mu$ that preserves the Ward identities 
of the electroweak gauge group}. This is exactly the choice we will proceed with. 

It is also worth noting that there is in fact a one-to-one correspondence between the 
Wess-Zumino counterterm (and, consequently $C_1 - C_N$ combination in Eqs.~\eqref{eq:trans1} 
and~\eqref{eq:trans2}) and the momentum shift $a$ that we are required to choose. If we choose the 
counterterm such that $C_N$ vanishes, and consequently, the effective action is invariant under the 
$SU(N)$ transformation (that we identify with the EW group transformation), we will not need 
any momentum shift between the two diagrams to restore gauge invariance. 
This is because the counterterm is imposing EW gauge invariance already. 
On the other hand, if we are not enforcing gauge invariance at the Lagrangian level 
with an appropriate Wess-Zumino counterterm, we are obliged 
to do it by choosing a non-trivial momentum shift, so that eventually all the three (and higher) point functions of the theory are well defined. In all of our further
calculations we will set the counterterm to zero and calculate the necessary momentum 
shift to restore gauge invariance. 

Finally, we briefly comment on the restoration of gauge invariance of the spontaneously broken 
gauge theory, that is the case in the EW group, and is relevant for the calculation of the 3-point function
of the $Z'$ with $W$ and $Z$ bosons. 
In principle, we use exactly the same procedure that we  have described before, except that when calculating the Ward identities, we have to include the piece  contributed by the Goldstone boson. 
Further details of this procedure are described in Ref.~\cite{Racioppi:2009yxa}. Practically, instead of Eq.~\eqref{eq:SimpleWard} we must demand
\beq
-i p_\mu \Gamma^{\mu \nu \rho} + m_V \Gamma^{\nu \rho}_\text{NG} = 0
\eeq 
where $\Gamma^{\nu \rho}_\text{NG}$ stands for the three-point function with the gauge boson $V_\mu$ replaced by its corresponding Goldstone boson.

\section{Dark Matter Models with Heavy Anomalous $Z'$ }
\label{sec:models}

In this section we describe in more detail the dark matter models 
with $Z'$ mediated interactions that we will consider.  
As we outline these models, we will make no requirement that the low-energy 
fermion content of our theory cancels all the gauge anomalies. 
This is a common step in the DM 
literature, 
which typically 
assumes that extra fermions, resolving anomaly cancellation, appear at high scales. 
Below the mass scale of these fermions we get an effective field theory similar to
the one that we have formulated in the previous section.

We begin with a SM singlet Majorana fermion $\chi$ that couples axially to the 
gauge boson $Z'$ 
of some new $U(1)'$ symmetry. 
The choice of this setup is mostly motivated by the null results of 
DM direct detection experiments. The vectorial couplings of the Majorana fermion to the 
$Z'$ are naturally precluded and therefore the scattering in the direct 
detection experiments
is either spin-dependent or velocity-suppressed
at tree level. The spin and velocity independent interactions are often negligible. 
Because the DM is not charged under the SM gauge groups, it has no impact 
on the mixed anomalies, in which we are mostly going to be interested in order to calculate
the DM annihilations into SM particles. 

While Majorana particles, being real fields, cannot be charged
under an \emph{exact} Abelian group, 
they can couple to
the gauge boson if the gauge group is broken.
In the latter case the fermions
get their masses via the Higgs mechanism (e.g. via couplings like $\sim \Phi \psi \psi$),
or, in the case of vector-like fermions, as a result of mixing with other singlet 
fermions. Because of the possible mixing effects, the coupling of $Z'$ to DM 
need not be equal to the coupling to the SM.

In general, the $U(1)'$ will be anomalous without the introduction of additional fermions 
besides $\chi$. Indeed, gauging any flavor-universal symmetry other than 
$B - L$, $Y$-sequential or linear 
combination thereof, leads to mixed anomalies between the gauge groups of the SM and the 
new $U(1)'$. These must  be resolved by new fermions with non-trivial SM charge. 
Here we do not try to build a full UV-complete model
(for explicit attempts to do this see e.g.~\cite{Ismail:2016tod,Ellis:2017tkh}), 
as for our purposes the only relevant quantities are the anomaly coefficients in the EFT 
that solely include the SM, $\chi$, and $Z'$. 
At sufficiently high energies, above the cutoff $\Lambda$ (see Sec.~\ref{sec:eftanomaly}),
all gauge anomalies must cancel.

As we are interested in the effects of anomalies, it is interesting to consider 
explicit models and discern the phenomenological importance of the effects of the $Z'$ couplings to 
the gauge bosons.
The first model we will be concerned with is  
one where the SM fermions are axially charged under $U(1)'$. 
In choosing this particular 
case we are mostly motivated by the vast existing literature on DM simplified models, that 
usually assumes a $U(1)'$ with pure axial charges as a standard benchmark 
point~\cite{Lebedev:2014bba,Boveia:2016mrp,Albert:2017onk}. 
This choice however comes with its own obvious shortcoming, that eventually renders it 
somewhat non-generic compared to the landscape of other options. 

For a SM fermion $f$ the usual SM Yukawa coupling $y_f H \bar{f} f$ is gauge invariant 
only if 
the Higgs doublet also has dark charge~\cite{Bell:2016uhg}. If $H$ is charged under $U(1)'$, 
in turn, then the $Z'$ acquires at least some mass from electroweak symmetry breaking and mixes 
with the $Z$. This $Z$-$Z'$ mixing is constrained by electroweak precision, and even though 
it can be viable if the $Z'$ mass is heavier than a few TeV~\cite{Jacques:2016dqz}, we 
prefer to avoid these 
complications, which would defocus us from the goal of showing the phenomenological impact of anomaly-induced interactions. If we assume that the SM Higgs is not 
charged under the $U(1)'$, the only option is to promote the Yukawa couplings to 
$U(1)'$ spurions, by writing the Yukawa terms as
\beq 
\left(\frac{\langle\Phi\rangle}{M_*}\right)^{2n} \tilde{y}_f H \bar{f} f~,
\eeq 
where $\langle\Phi\rangle$ is the vacuum expectation value of a field $\Phi$ that 
spontaneously breaks  $U(1)'$, 
$M_*$ is some suppression scale dictated by the UV completion, and 
$n$ is the ratio of the fermion axial $U(1)'$ charge to the $\Phi$ charge. In this framework, 
the natural size of the Yukawa couplings is driven by the size of 
$\langle\Phi\rangle/M_*$. 
Although this approach is generally consistent with the smallness of the SM
Yukawas, it becomes difficult to reproduce the top Yukawa coupling in this way.   
To ``fix'' this problem we assume that the top quark couples \emph{vectorially} to the $Z'$. 
The other fermions are taken to have axial couplings, except for the neutrinos which 
necessarily have purely left-handed couplings. We will further call this particular symmetry
$\Upaxt$.

As with the scalar that could be responsible for a DM Majorana mass term, 
the particular characteristics of the scalars that generate SM fermion Yukawas are not 
relevant to the interactions at hand. For our purposes, the only other effect of the 
scalars which acquire 
$U(1)'$-breaking vacuum expectation values is to provide mass to the 
$Z'$.\footnote{For more comments on the possible relevant effects of the $U(1)'$-breaking
Higgs see~\cite{Kahlhoefer:2015bea}.}
We simply parametrize these effects by a mass term $\frac{1}{2} \mZp^2 (Z')^2$, and 
generally ignore the details of the scalar sector from here on.\footnote{As we have mentioned, another 
possibility to deal with the fermion masses problem 
would be to charge the SM Higgs under the $U(1)'$ and deal with the $Z-Z'$
mixing similarly to~\cite{Jacques:2016dqz}. See also~\cite{Cui:2017juz} for some important insights 
on this framework.}

In order to summarize these considerations and to fix our notation, we show here the newly added terms to the Lagrangian:
\begin{multline}
\mathcal L_\text{DM} = 
        -\frac 14 Z'_{\mu\nu} Z^{\prime\, \mu\nu} 
        +\frac 12 \mZp^2 Z^{\prime\, 2}_\mu 
        + \frac 12 \overline \chi (i \slashed \partial  -\mX) \chi 
        \\
        + \frac 12 \gX Z^{\prime \, \mu} \overline \chi \gamma_\mu \gamma_5 \chi 
        + \gZp Z'_\mu \sum_f \overline f \, \left(g_V^f \gamma^\mu+g_A^f \gamma^\mu \gamma_5\right)\, f  \,,
\end{multline}
where the coupling of the $Z'$ to the SM fermions $f$ is given by $\gZp$ times the charges $g_V^f$ and $g_A^f$, which are given in Table~\ref{tab:U1 charges}, and $\gX$ is the coupling to the Majorana DM $\chi$.

It is easy to note in what sense this ``modified axial model'' $\Upaxt$ is not generic. 
If we do not tighten the solution of the flavor problem with the DM theory (which 
is possible but by no means necessary) yet still insist that the SM Higgs is uncharged 
under the new force, the charges of the SM fermions are vectorlike under the 
$U(1)'$. This immediately implies that the mixed anomalies with the $\Uem$
and the SM color group must vanish. At sufficiently large DM masses this 
strongly suppresses the $\gamma \gamma$ and $gg$ annihilation channels of the DM, 
but does not qualitatively change other channels. As an example of this model 
we choose $\UBpL$, which is simply one representative point in a 
large class of models. 

Finally we choose to also show a leptophilic model (for this purpose, $\UL$).
This choice is special because we have no constraints from the LHC and direct 
detection, and all the constraints come from indirect detection 
searches.\footnote{Strictly speaking, this model is not totally invisible to 
direct detection experiments due to radiative couplings to the hadrons 
(for works along these lines see~\cite{Kopp:2014tsa}). 
However the effect is expected to be so small that we disregard it here.}

\begin{table}[t]
\centering
\begin{tabular}{cccccccc}
\toprule
 & $SU(3)$ & $SU(2)$ & $\UY$  & $\UBL$  & $\Upaxt$ & $\UBpL$ & $\UL$\\
\midrule
$\begin{pmatrix} \nu^{e}_L \\ e^i_L \end{pmatrix}$, 
$\begin{pmatrix} \nu^{\mu}_L \\ \mu^i_L \end{pmatrix}$, 
$\begin{pmatrix} \nu^{\tau}_L \\ \tau^i_L \end{pmatrix}$
& $\textbf{1}$ & $\textbf{2}$ & $-\tfrac 12$ & $-1$ & $-1$ & $+1$ & $+1$  \smallskip\\
$\left(e^i_R\right)^\text{C}$,
$\left(\mu^i_R\right)^\text{C}$,
$\left(\tau^i_R\right)^\text{C}$
 & $\textbf{1}$ & $\textbf{1}$ & $1$ & $+1$ & $-1$ & $-1$ & $-1$ \smallskip\\
 \midrule
$\begin{pmatrix} u_L\\ d_L\end{pmatrix}$,
$\begin{pmatrix} c_L\\ s_L\end{pmatrix}$
& $\textbf{3}$ & $\textbf{2}$ & $\tfrac 16$ & $+\tfrac 13$ & $-1$ & $+\tfrac13$ & $0$
\smallskip\\
$\left( u_R\right)^\text{C}$, 
$\left( c_R\right)^\text{C}$ 
& $\overline{\textbf{3}}$ & $\textbf{1}$ & $-\tfrac 23$ & $-\tfrac 13$ & $-1$  & 
$-\frac13$ & $0$ \smallskip\\
$\left( d_R\right)^\text{C}$,
$\left( s_R\right)^\text{C}$
& $\overline{\textbf{3}}$ & $\textbf{1}$ & $\tfrac 13$ & $-\tfrac 13$ & $-1$  & $-\tfrac13$ 
& $0$
\smallskip\\
 \midrule
$\begin{pmatrix} t_L\\ b_L\end{pmatrix}$ & $\textbf{3}$ & $\textbf{2}$ & $\tfrac 16$ & $+\tfrac 13$ & $-1$ & $+\tfrac13$ & $0$ \smallskip\\
$\left( t_R\right)^\text{C}$ 
& $\overline{\textbf{3}}$ & $\textbf{1}$ & $-\tfrac 23$ & $-\tfrac 13$ & $+1$  & $-\tfrac13$ 
& $0$
\smallskip\\
$\left( b_R\right)^\text{C}$
& $\overline{\textbf{3}}$ & $\textbf{1}$ & $\tfrac 13$ & $-\tfrac 13$ & $-1$  & 
$-\tfrac13$ & $0$ 
\smallskip\\
 \midrule
Higgs $\begin{pmatrix} \phi^+ \\ \phi^0\end{pmatrix}$ & $\textbf{1}$ & $\textbf{2}$ & $\frac 12$ & $0$ & $0$ & $0$ & $0$ \smallskip\\
\bottomrule
\end{tabular}
\caption{Charges of the SM matter content under some choices of $U(1)'$ that we further analyze 
in the paper. }
\label{tab:U1 charges}
\end{table}

Taking all this into account we present the charges of the SM fields under the new 
$U(1)'$s in Table~\ref{tab:U1 charges}. 
For comparison, we also show $B - L$, 
which is of course anomaly free and does not require any extra terms in the 
effective action.

Since our axial vector 
model features flavor non-universal $Z'$ quark couplings, 
we parenthetically consider here the flavor constraints on this kind of $Z'$. 
Even though the axial-$Z'$ couplings are diagonal in the flavor basis, the quark rotations 
that diagonalize the Yukawa matrix generally induce off-diagonal couplings between quark 
mass eigenstates~\cite{Abdallah:2015ter}. 
To estimate the size of the associated flavor-changing neutral currents (FCNCs), 
we must make assumptions about the structure of the quark rotations. 
Note that the only measured misalignment between the quark flavor and 
mass eigenstates is from the CKM matrix $V_{\mathrm{CKM}}$, which is a combination of 
the two
left-handed quark rotations. Conversely, the $\Upaxt$ model only 
contains non-universality in the right-handed up-quark sector.
Of course, if the mixing angles in the RH sectors are completely anarchical, 
the structure that we discuss is not viable. However, this is not the only option, 
especially if we take into account the hierarchical structure of $ V_\mathrm{CKM}$.   
First, FCNCs may be completely avoided if the right-handed quark flavor and 
mass eigenstates are identical (this would invoke either a fine-tuning
or some other structure that would explain the vanishing rotation angles). 

Alternatively, let us assume that the flavor structures of the RH and LH quark sectors are similar, such that the product of the rotations between the up- and down-type RH quark U(1)' flavor and mass eigenstates is $\sim V_\mathrm{CKM}$.
Then, since the non-universality is only in the third generation, the $Z' \bar{c}_R u_R$ 
coupling will go as $\sim V_{\mathrm{CKM}}^{ub*} V_{\mathrm{CKM}}^{cb}$, 
which is quite small, without dangerous consequences for $D$ mixing. 
Non-universal couplings in the down-type sector are also induced at the loop level, 
leading to effects such as $B-\overline B$ mixing.

Finally we note that a kinetic mixing 
term $B_{\mu\nu} F'^{\mu\nu}$, where $B$ and $F$ are the $U(1)_Y$ and $U(1)'$ field 
strengths respectively, is fully allowed by the symmetries of the theory. 
Sizable kinetic 
mixing can lead to observable effects that are interesting but separate from those caused by 
the triple gauge vertices induced by anomalies. We henceforth assume negligible mixing,
 and concentrate on the anomalous couplings among the SM and $U(1)'$ gauge bosons. 

\section{Application to Dark Matter Models}
\label{sec:dm}

In this section we present the main results of our paper. First, we 
will use the results of Sec.~\ref{sec:eftanomaly} to explicitly calculate the 
annihilation cross sections of the DM particle into the SM gauge bosons.
In the following subsections we show the prospects for the direct and indirect detection, 
as well as LHC searches. We will emphasize the complementarity of these searches to 
properly analyze the possible parameter space of these models. 

\subsection{Annihilation cross sections into the SM gauge bosons}
The objective in this part of our paper is to explicitly calculate the relevant 
annihilation cross sections that arise only at the one-loop level. 
To begin, we outline the calculation of the coupling between three gauge bosons induced by anomalies, starting with the $Z'$-$\gamma$-$\gamma$ vertex. We take a single fermion 
$f$ of electric charge $Q_f^{em}$ to run in the loop diagrams of 
Fig.~\ref{fig:VertexFunction}, whose amplitude we write as 
$\epsilon_\mu(p_1) \epsilon^*_\nu(p_2) \epsilon^*_\rho(p_3) \Gamma^{\mu\nu\rho}$. 
Note that we use $p_1$ for the $Z'$ momentum, while the momenta $p_2,\ p_3$ stand for the photon momenta.
If the fermion's $U(1)'$ coupling is vectorial, then by Furry's theorem the vertex vanishes. 
Without loss of generality we assume a $Z'f\bar{f}$ vertex with strength 
$i \gZp \gamma^\mu (g_V + g_A \gamma^5)$ with the understanding that only $g_A$ will 
contribute.\footnote{Note that this parametrization is completely generic and suitable for analyzing any anomalous $Z'$. If we turn back to the models we have outlined 
in Sec.~\ref{sec:models}, we see that in those particular models all the SM fermions have 
either $g_V = 0$ or $g_A = 0$, but this is by no mean guaranteed for a generic $Z'$.} 
As described in Sec.~\ref{sec:eftanomaly}, contracting the external gauge boson momenta 
with the triangle amplitude gives non-vanishing results due to surface terms 
(see~\cite{Preskill:1983,Racioppi:2009yxa} for further calculation 
details).\footnote{{Notice that the Landau-Yang theorem forbids resonant production when the $Z'$ is on shell; for further details, see \cite{Maltoni}.
See also \cite{Garcia-Cely:2016hsk} for a discussion about this calculation within the Landau gauge.}}
 The resulting Ward identities depend, as explained in Sec.~\ref{sec:eftanomaly}, on the 
loop momentum shift $a$:
\beqa
\label{eq:zpyywi}
(p_1)_\mu \Gamma^{\mu\nu\rho} &=& \frac{\gZp e^2 g_A (Q_f^{em})^2}{8\pi^2} \epsilon^{\nu\rho\alpha\beta} a_\alpha (p_1)_\beta \nonumber \\
(p_2)_\nu \Gamma^{\mu\nu\rho} &=& \frac{\gZp e^2 g_A (Q_f^{em})^2}{8\pi^2} \epsilon^{\mu\rho\alpha\beta} (a + 2 p_3)_\alpha (p_2)_\beta \\
(p_3)_\rho \Gamma^{\mu\nu\rho} &=& \frac{\gZp e^2 g_A (Q_f^{em})^2}{8\pi^2} \epsilon^{\mu\nu\alpha\beta} (a - 2 p_2)_\alpha (p_3)_\beta \nonumber
\eeqa

At this stage we can either tune the Wess-Zumino conterterm to get rid of any $a_\mu$ dependence 
in these expressions or, alternatively, set the Wess-Zumino counterterm to zero and find an
appropriate momentum shift to maintain the necessary Ward identities. We choose the latter 
recipe to resolve this problem. 
We make the phenomenologically motivated choice of retaining $\Uem$ gauge invariance, 
which corresponds to the requirement that the second and third lines of Eq.~\eqref{eq:zpyywi} 
vanish. This, in turn, may be accomplished by setting $a = 2 (p_2 - p_3)$, yielding the Ward identities
\beqa
\label{eq:zpyywi2}
(p_1)_\mu \Gamma^{\mu\nu\rho} &=& \frac{g' e^2 Q_f (Q_f^{em})^2}{2\pi^2} 
\epsilon^{\nu\rho\alpha\beta} (p_2)_\alpha (p_3)_\beta \nonumber \\
(p_2)_\nu \Gamma^{\mu\nu\rho} &=& (p_3)_\rho \Gamma^{\mu\nu\rho} = 0
\eeqa

Next, to calculate the relevant cross section, 
we write the most general form of the amplitude using the standard 
Rosenberg parametrization~\cite{Rosenberg:1962pp}
\beqa
\label{eq:rosenberg}
\Gamma^{\mu\nu\rho} &=& \frac{\gZp e^2 g_A (Q_f^{em})^2}{\pi^2} \Big( I_1 \epsilon^{\alpha\nu\rho\mu} (p_2)_\alpha + I_2 \epsilon^{\alpha\nu\rho\mu} (p_3)_\alpha \nonumber \\
&&\ +\ I_3 \epsilon^{\alpha\beta\nu\mu} (p_2)^\rho (p_2)_\alpha (p_3)_\beta + I_4 \epsilon^{\alpha\beta\nu\mu} (p_3)^\rho (p_2)_\alpha (p_3)_\beta \\
&&\ +\ I_5 \epsilon^{\alpha\beta\rho\mu} (p_2)^\nu (p_2)_\alpha (p_3)_\beta + I_6 \epsilon^{\alpha\beta\rho\mu} (p_3)^\nu (p_2)_\alpha (p_3)_\beta \Big) \nonumber
\eeqa
where $I_i, 1 \leq i \leq 6$ are form factors to be computed. By dimensional analysis, it is clear that the effect of any divergences must be in $I_1$ and $I_2$, while the remaining form factors are finite. We thus use the Ward identities of Eq.~\eqref{eq:zpyywi2} 
to fix the divergent form factors, and calculate the others explicitly. 
The final result is~\cite{Racioppi:2009yxa}
\beqa
\label{eq:formfactors}
I_1(p_2, p_3; m_f) &=& (p_2 \cdot p_3) I_3(p_2, p_3; m_f) + p_3^2 I_4(p_2, p_3; m_f) \nonumber \\
I_2(p_2, p_3; m_f) &=& -I_1(p_3, p_2; m_f) \nonumber \\
I_3(p_2, p_3; m_f) &=& -C_{12}(p_3^2, p_1^2, p_2^2, m_f^2, m_f^2, m_f^2) \\
I_4(p_2, p_3; m_f) &=& C_{11}(p_3^2, p_1^2, p_2^2, m_f^2, m_f^2, m_f^2) + C_1(p_3^2, p_1^2, p_2^2, m_f^2, m_f^2, m_f^2) \nonumber \\
I_5(p_2, p_3; m_f) &=& -I_4(p_3, p_2; m_f) \nonumber \\
I_6(p_2, p_3; m_f) &=& -I_3(p_2, p_3; m_f) \nonumber
\eeqa
where the $C$ functions are Passarino-Veltman loop functions~\cite{Hahn:1998yk}. When there are multiple fermions charged under both electromagnetism and $U(1)'$, Eq.~\eqref{eq:rosenberg} 
is readily generalized by summing over the available loop fermions.

The above vertex may now be used to calculate physical observables. For instance, the amplitude for DM annihilation to photons immediately follows, and the resulting cross section takes a 
rather compact form\footnote{Hereafter we use the complex mass scheme to treat the cross sections 
near the resonances. } 
\begin{small}
\beq
\label{eq: xsec gamma gamma}
\sigma(\chi\chi \to \gamma\gamma) =
\frac{\aem^2 \gX^2 \gZp^4}{\pi^3}
\frac{\mX^2 \sqrt{s}}{ \mZp^4 \sqrt{s-4\mX^2} } 
	\times \Bigg|\sum_f \cAf N_c^f  Q_f^2
	\Big[
		2 m_f^2 C_0(0,0,s,m_f^2,m_f^2,m_f^2) + 1
	\Big] \Bigg|^2 \,,
\eeq
\end{small}
where the explicit form of the Passarino-Veltman function involved is
\beq
C_0(0,0,s,m_f^2,m_f^2,m_f^2) = \frac{1}{2s}\log ^2\Bigg(\frac{\sqrt{s (s - 4 m_f^2)}+2 m_f^2-s}{2m_f^2}\Bigg) \,.
\eeq
This should be compared to the cross section for DM annihilation to fermions,
\begin{small}
\begin{multline}
\label{eq: xsec f f}
\sigma(\chi\chi \to f \bar f) = \frac{\gX^2 \gZp^4 \, N_c^f}{3 \pi s 
\big((s - \mZp^2)^2+ \GZp^2 \mZp^2\big)}
\sqrt{\frac{s - 4 \mf^2}{s - 4 \mX^2}} \Bigg(
g_V^2  (s - 4 \mX^2 ) (s+ 2 \mf^2 ) \\
\ +\ g_A^2 \bigg( s (s - 4 \mX^2)  + 4 \mf^2 
\Big( \mX^2 \Big(7  - 6 \frac{s}{\mZp^2} + 3 \frac{s^2}{\mZp^4}\Big) - s \Big)
 \bigg)
\Bigg) \,.
\end{multline}
\end{small}

The key difference between these cross sections is that the annihilation to photons 
remains constant with increasing center-of-mass energy, unlike the annihilation to fermions which 
eventually falls as $1/s$. Of course this ``constant'' annihilation rate cannot 
proceed to arbitrarily 
high energy because it would eventually break the unitarity of the theory. We will comment 
on this issue later in the section.

We calculate the form factors for the annihilations into the rest of the gauge bosons, using exactly 
the machinery that we have shown here. We list the relevant results in Appendix~\ref{sec:results}.
For simplicity we take the $Z'$ width to be $\Gamma_{Z'} = m_{Z'} / 10$ throughout our calculations. This choice only affects the extent of the influence of resonant effects in our results.

\begin{figure}[t]
\centering
\includegraphics[width=\textwidth]{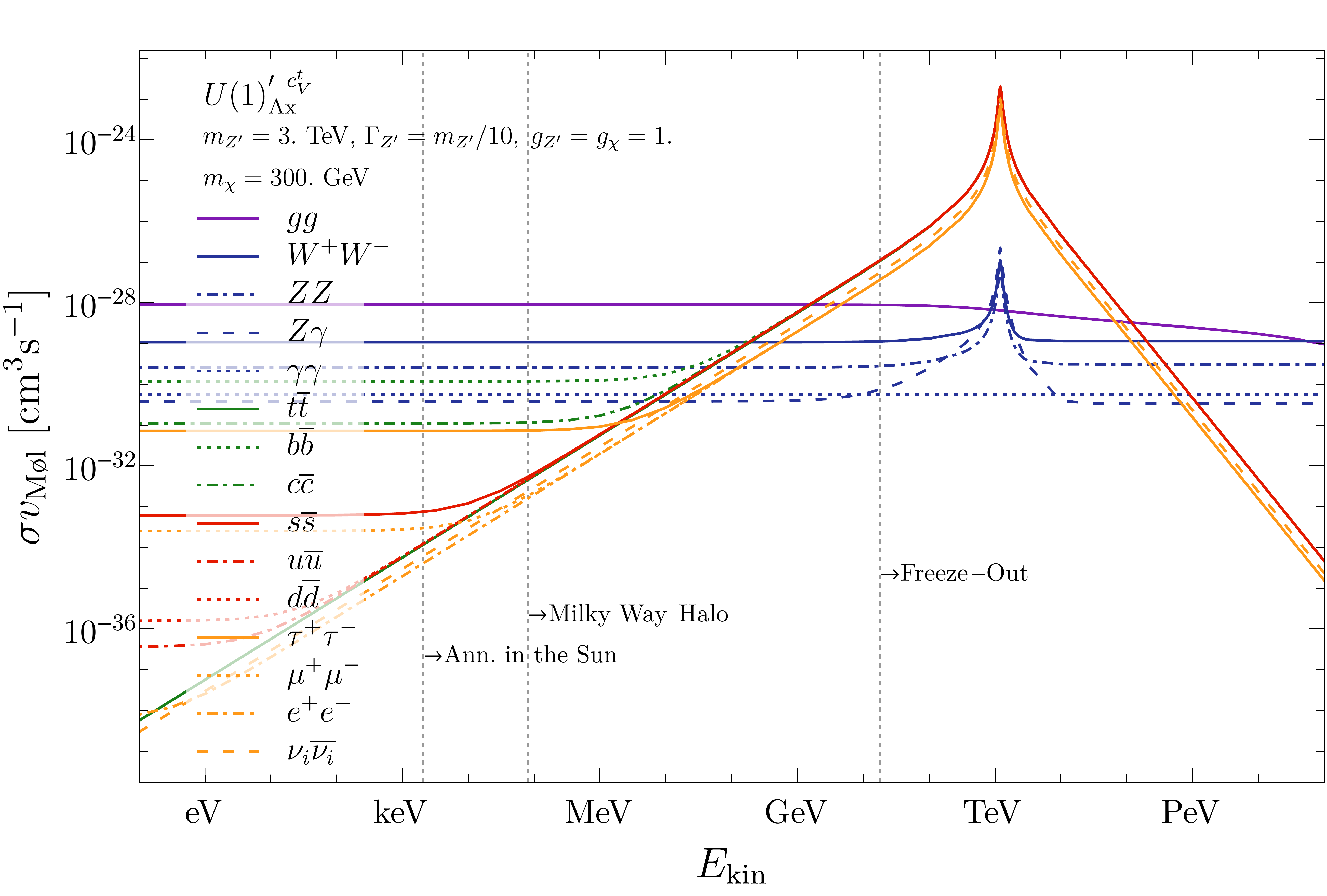}
\caption{\label{fig:sigmavek} Annihilation cross sections in the $\Upaxt$ model as a 
function of the DM kinetic energy within the EFT that we describe. 
The blue curves indicate gauge boson final states that receive contributions from anomalies. Annihilations into heavy quarks, light quarks, and leptons are shown in green, red, and yellow, respectively. Kinetic energies corresponding to DM in the Sun, 
the Milky Way halo, and at freezeout are indicated. Note that above the scale $\sim 100$~TeV 
the EFT cannot give the correct solution due to the inevitable unitarity breakdown. }
\end{figure}

Before we present our results, it is instructive to see how the 
annihilation cross sections $\sigma v$ scale with the kinetic energy of the fermions for 
fixed DM mass. We show this scaling 
\emph{within the EFT} on Fig.~\ref{fig:sigmavek} (we use for this illustration
the $\Upaxt$ model). While the cross sections into the fermions fall at the high energies 
as $1/s$ (as one would expect), the annihilations into the gauge bosons stay 
constant as a function of $s$, signaling an inevitable breakdown of unitarity at high 
energies. This breakdown is expected from the way we have formulated our EFT in 
Sec.~\ref{sec:eftanomaly}, in particular because of the higher dimensional interactions 
that we were forced to introduce. Of course these cross sections are tamed at the scale 
where the spectator fermions show up. This can in turn happen at or below the scale 
$\Lambda$ as defined in Eq.~\eqref{eq:cutoff} (modulo replacing the Abelian anomaly by 
a mixed one). 

Note also that unitarity will often dominate the exact bound on the cutoff 
$\Lambda$, although it will often be of order~\eqref{eq:cutoff}. For example, 
a simple back-of-the-envelope estimation leads us to the conclusion that the unitarity 
of the model depicted on Fig.~\ref{fig:sigmavek} \emph{will break down }  at a scale 
$\sim 100$~TeV. In this sense the very right corner of this plot is not meaningful 
and the physics there should be described by a full UV complete theory rather than the EFT.

Note also the difference between the fermions that couple axially and ones that couple 
vectorially to the $Z'$. While annihilations into the former final states (in this particular
example, all the SM fermions except the top) are constant at low energies, the latter in 
this range scale as $v^2$, and therefore linearly with the kinetic energy. This can also 
be clearly observed in Eq.~\eqref{eq: xsec f f}. 

Another important lesson that we learn from Fig.~\ref{fig:sigmavek} is the dominance of the various channels in different physical situations. For example, the velocity is still high enough during the 
thermal freeze-out to render the annihilation into the gauge bosons unimportant, such that 
the relic abundance is determined almost completely by the annihilations into the fermions.
However at lower velocities (annihilation in the Galactic halo or at the center of the Sun)
the entire signal is essentially determined by the radiative annihilations into the 
gauge bosons.    

Finally let us notice, that even in the models where the respective mixed anomaly 
vanishes, the annihilation channels into the gauge bosons are induced by finite radiative 
corrections.
However, because these contributions 
do not grow with energy, they are much smaller than the anomaly-augmented annihilations, and 
can be neglected.

\subsection{Relic abundance}

We first briefly comment on the DM relic abundance, if we assume that the DM is the thermal 
relic (which might or might not be the case). 
WIMP freezeout typically happens near $x = m_\chi / T \approx 25$, with the 
particular decoupling temperature only logarithmically sensitive to the annihilation cross 
section. The annihilation channels which determine the relic abundance are thus simply 
the modes which dominate at a DM velocity of $\sqrt{\frac{3}{x}} \sim 1/3$. 
From Fig.~\ref{fig:sigmavek}, it is apparent that DM annihilation to fermions is 
primarily responsible for setting the relic abundance. Consequently, the impact of anomalies 
in the DM relic density calculation is minimal. We thus expect the values of the couplings and masses that reproduce the observed DM abundance to be similar to 
previous calculations in the literature, see e.g.~\cite{Arcadi:2017kky}.

\subsection{Indirect detection}

Today very little kinetic energy is available for DM annihilation because the 
typical velocity of a DM in the Milky Way halo is $\sim 10^{-3}$. In our models
the gauge boson modes can dominate the annihilations, 
and so the DM can be probed through searches for annihilation to 
 $g g$, $W^{+} W^{-}$, $\gamma \gamma$, $Z \gamma$ and $ZZ$. 

\begin{figure}[t]
\centering
\includegraphics[width=0.49\textwidth]{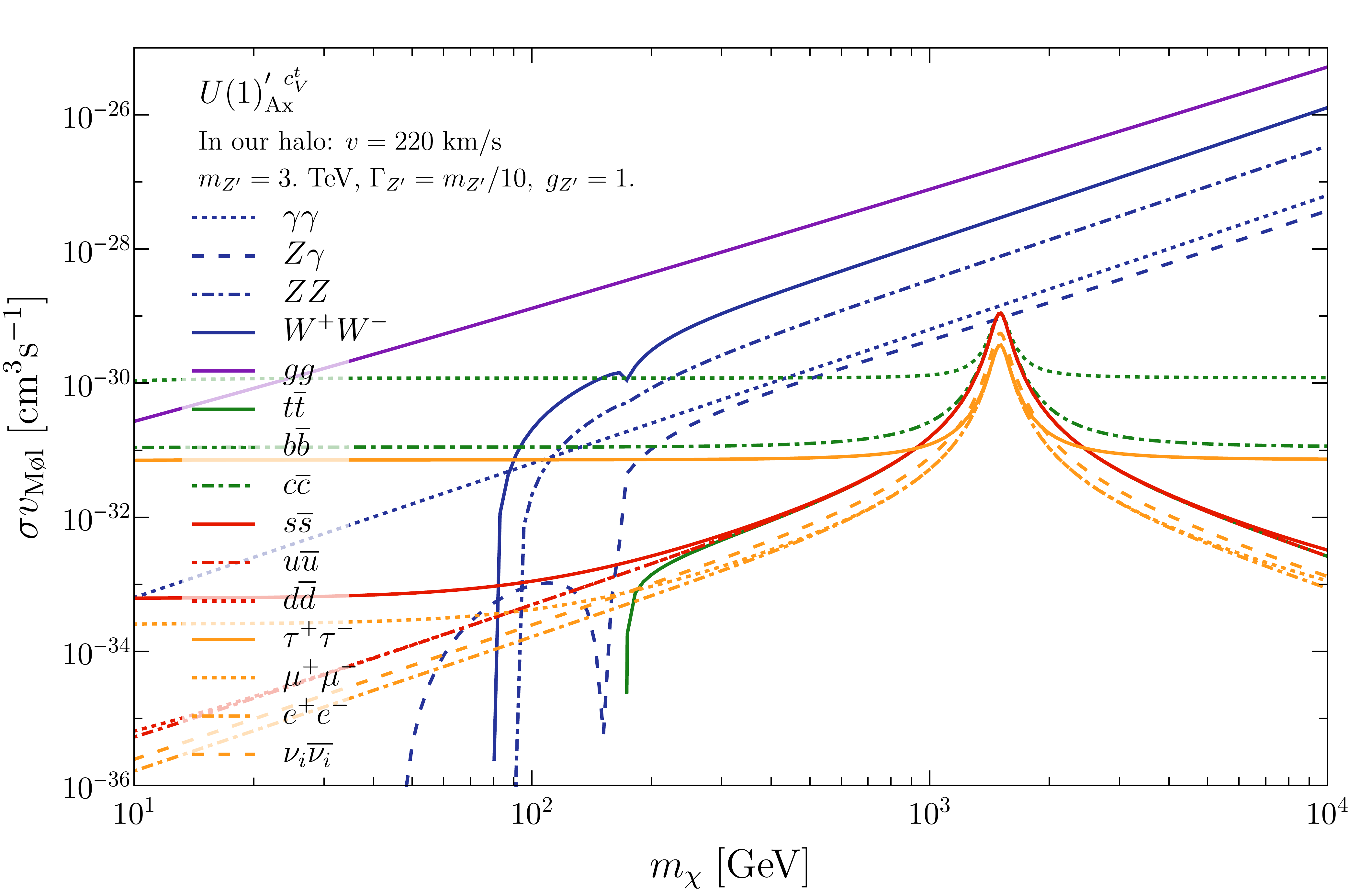}
\hfill
\includegraphics[width=0.49\textwidth]{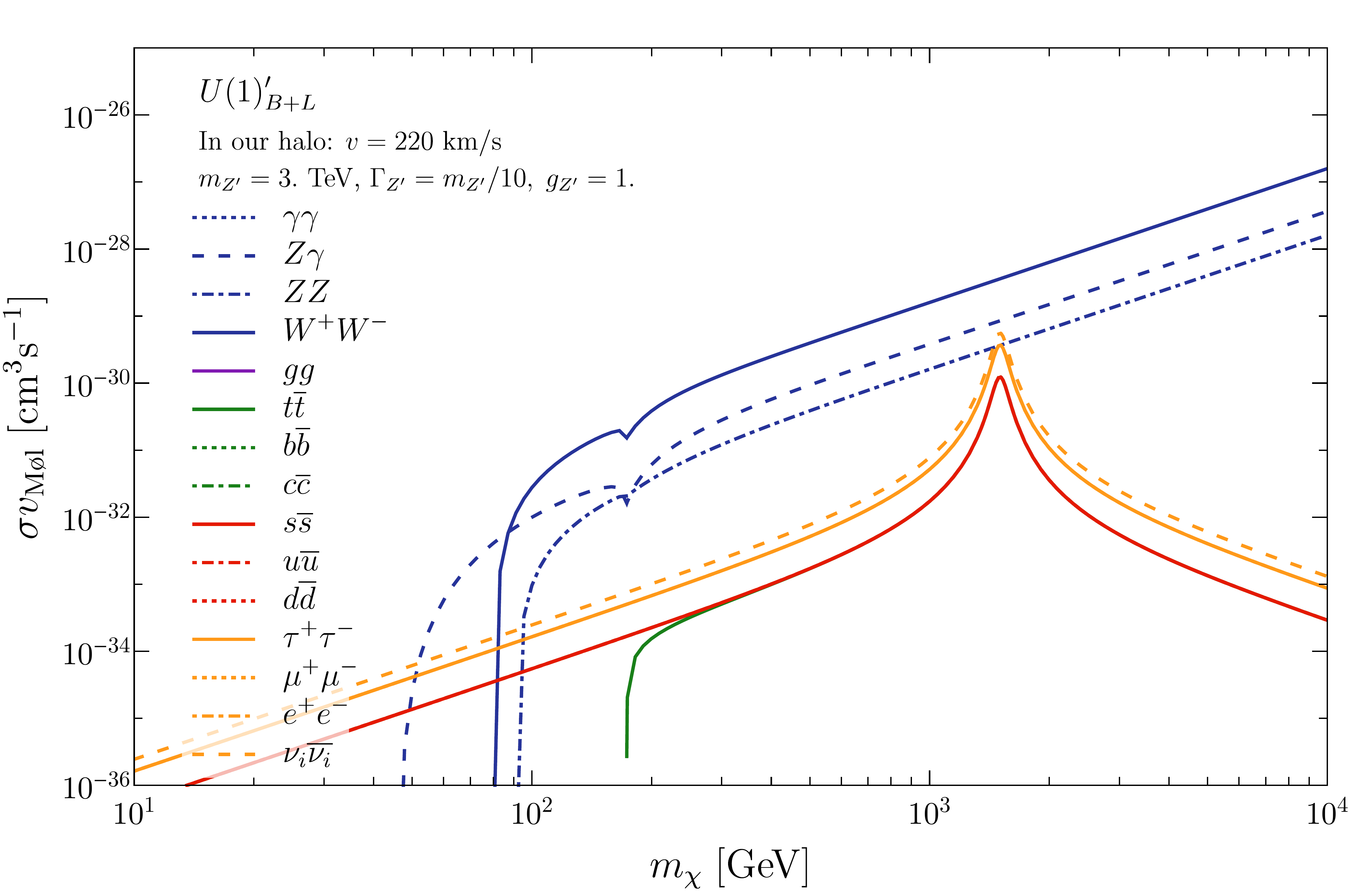}
\caption{\label{fig:brhalo} Left: DM mass vs. annihilation branching ratios in the $\Upaxt$ model.
The DM velocity is taken to be 220~km/s, characteristic of the Milky Way halo. 
The curves are colored as in Fig.~\ref{fig:sigmavek}. Right: same for the $\UBpL$ model. 
The quark lines all overlap on the red line. The branching ratios of the $\UL$ model 
are analogous to the 
ones of $\UBpL$, without the quark channels.}
\end{figure}

 We illustrate this point on Fig.~\ref{fig:brhalo}, where we show the 
 branching ratios into the various annihilation channels of the DM in our galaxy 
 as a function 
 of the DM mass. 
 If the DM is relatively light, $\sim 10$~GeV, 
 the BRs are dominated by the fermionic
 channels, particularly $b \bar b$.  
 However at sufficiently high DM masses the $gg$ 
 (if the mixed anomaly of the $U(1)'$ with the SM does not vanish) 
 and $W^+W^-$ channels 
 dominate the annihilations at such low DM velocities, 
 and therefore the indirect detection signatures. 
 We also point out  the  importance of the  
 $\gamma \gamma$ (when present) and $Z \gamma$ annihilation channels. 
 Although the latter channels  are heavily suppressed compared to the 
 $W^+W^-$, the photon emission is monochromatic, leading to the prediction of a $\gamma$ ray line.

\subsubsection{Gamma ray continuum searches}
We first consider limits from the continuum $\gamma$ ray spectrum, 
where the strongest current bound comes from dSph Fermi-LAT 
observations~\cite{Drlica-Wagner:2015xua,Ackermann:2015zua} 
and, for TeV scale DM masses, from HESS observation 
of the continuum emission from the Galactic Center~\cite{Abdallah:2016ygi}.

 These bounds depend on the products of DM annihilation, 
 as different SM particles yield distinct photon spectra. 
 In order to apply the $\gamma$ ray limits, we thus consider 
 the annihilation branching ratios to different final states. 
 At low DM mass, the fermionic annihilation channels are dominant, 
 as seen in Fig.~\ref{fig:brhalo}. 
 
 We start by discussing the $\Upaxt$ model, where all such channels except for 
 $t\bar{t}$ are chirally suppressed, and the $b\bar{b}$ 
 and $\tau^+ \tau^-$ 
 annihilations are more common than those into the light fermions.  
 In practice, the limits on annihilations to $b\bar{b}$ 
 and $\tau^+ \tau^-$ are quite close to one another~\cite{Ackermann:2015zua}. 
 Similarly, DM annihilations to charm quarks produce similar photon spectra 
 as to up quarks~\cite{Cirelli:2010xx}, for which the limits are in turn close 
 to those for annihilations to $b\bar{b}$. We thus choose to compare the 
 total fermionic annihilation cross section to the Fermi-LAT limit on DM annihilating 
 to bottom quarks for DM masses below approximately 200~GeV.

 At larger DM masses, the $gg$, $W^+ W^-$ and $ZZ$ channels take over. 
 Again, since the resulting $\gamma$ ray spectra from these annihilation modes are 
 similar, we simply compare the total bosonic annihilation cross section to the 
 Fermi-LAT limit on DM annihilations to $W^+ W^-$ 
 (which gives a slightly weaker bound than $gg$).
 Finally, in the resonance region $m_\chi \approx m_{Z'} / 2$, 
 fermionic annihilations take over again 
 and we switch back to comparing the total annihilation cross section to the $b\bar{b}$ limit from Fermi-LAT once more. Throughout, we assume that $\chi$ makes up all of the observed 
 DM.\footnote{The dip in the gauge boson BRs on the resonance is due to the Landau-Yang 
 theorem, which holds approximately even for the EW bosons because $m_Z \ll m_{Z'}$.}
 
 In the $\UBpL$ and $\UL$ models the procedure is analogous, with the notable difference 
 that the $gg$ channel disappears. At low DM masses the annihilations to fermions, 
 which now couple vectorially to the $Z'$, are velocity suppressed.

\begin{figure}[tbp]
\centering
\includegraphics[width=0.5\textwidth]{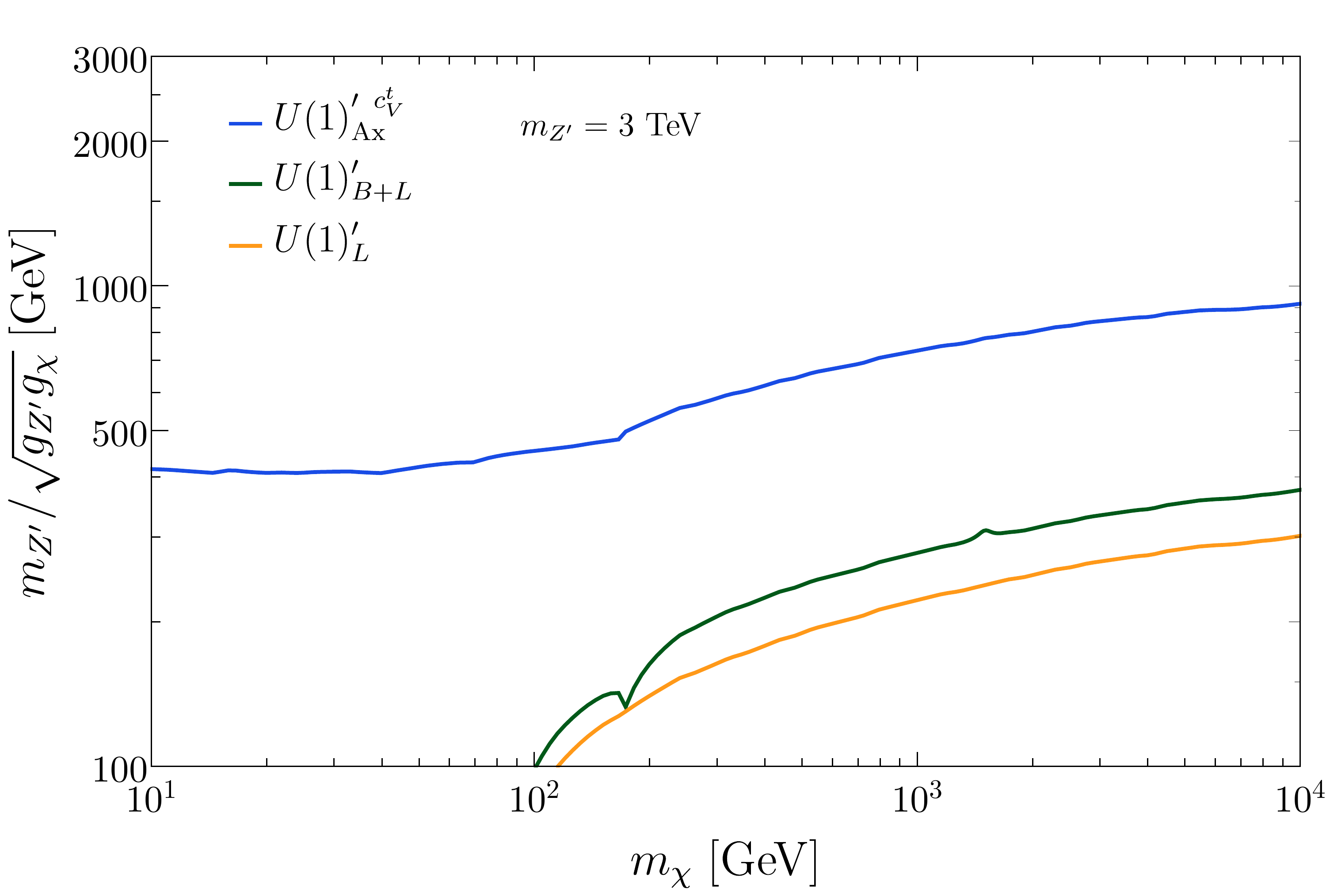}\hfill 
\includegraphics[width=0.5\textwidth]{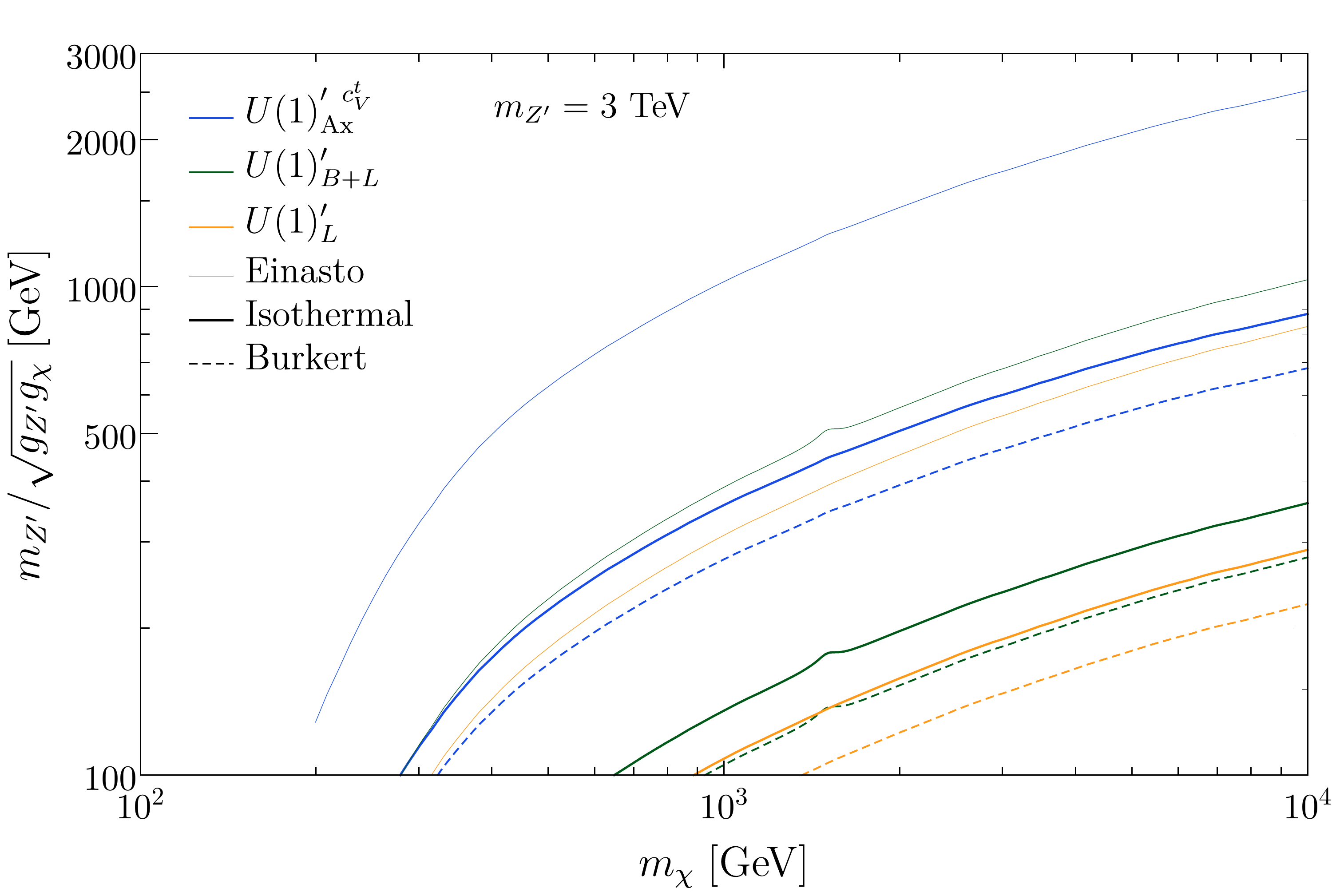}
\caption{\label{fig:continuum limits} Limits from continuum $\gamma$ ray emission on the three models we consider. 
\textit{Left}:
bounds from Fermi-LAT observations of dSphs~\protect{\cite{Drlica-Wagner:2015xua,Ackermann:2015zua}}.
\textit{Right}: bounds from HESS observation of the Galactic Center~\protect{\cite{Abdallah:2016ygi}}, for three choices of the DM profile distribution.
}
\end{figure}

We show the bounds on the suppression scale  $m_{Z'} / \sqrt{\gZp g_\chi}$ in the three models 
as a function of the DM mass in Fig.~\ref{fig:continuum limits}.
These bounds are insensitive to the choice of DM profile~\cite{Ackermann:2015zua}. 
The bound on the $\Upaxt $ model is significantly stronger than those on the $B+L$ and $L$ models 
because of the dominance of the $gg$ annihilation channel, which is prolific in $\gamma$ rays due 
to its secondary production of pions. As the mixed anomaly of the latter two $U(1)$s with the color 
group vanishes, the annihilations into $gg$ in these models are much more modest. 

The choice of the mediator mass has no effect on the bounds except in the resonance region, 
and so while the limits correspond to somewhat large couplings for the indicated $m_{Z'}$. 
Lighter mediators will have similar constraints. 
Note that if the DM is significantly heavier than the mediator mass, the coupling $g_\chi$
should be sufficiently small to avoid unitarity constraints on the DM 
self-scattering~\cite{Kahlhoefer:2015bea}.

We show the HESS continuum Galactic Center bounds in Fig.~\ref{fig:continuum limits}, assuming
three different DM profiles
(see Sec.~\ref{sec:gamma line searches} for more details). For each profile we compute the 
integrated J-factor between 0.3\textdegree\ and 1\textdegree\ around the direction 
of the Galactic Center using the tables from~\cite{Cirelli:2010xx} and scale the HESS bound appropriately.

\subsubsection{Gamma ray line searches}
\label{sec:gamma line searches}
Given the potential for annihilation to $\gamma \gamma$ or $Z \gamma$ through anomalies, we now discuss the impact of $\gamma$ ray line searches on our benchmark models, as performed by Fermi-LAT~\protect{\cite{Ackermann:2015lka}} and HESS~\protect{\cite{Abramowski:2013ax}}. Fermi-LAT is typically sensitive to photons below several hundred GeV in energy, while HESS, 
being a terrestrial telescope, has the best sensitivity for  
much more energetic $\gamma$ rays. 

The bounds from line searches generally depend on the DM halo profile, 
and so we will show their variation when different profiles are considered;
for an overview of DM halo profiles see for instance Ref.~\cite{Lisanti:2016jxe}.
Fermi-LAT optimizes the signal region of interest to maximize the bound depending on the 
profile, for several different halo shape choices. For instance, the optimal bound is 
obtained for a region subtending 16\textdegree\ around the galactic center for the 
Einasto profile, but 90\textdegree\ for an isothermal profile. HESS only shows limits 
for the Einasto profile, using a signal region of radius 1\textdegree. 
We choose to show bounds for Einasto, isothermal and Burkert DM halo profiles, 
by rescaling the Fermi-LAT and HESS limits using the ratios of J-factors for 
different profiles over the signal regions of interest~\cite{Cirelli:2010xx}. In the case of Fermi-LAT, we obtain the limit for a Burkert profile by rescaling the constraint for an isothermal profile, as these halo shapes are both relatively cored.

We further calculate the expected annihilation cross section to photons and compare with the Fermi-LAT and HESS $\gamma$ ray line search bounds, computed as described above for different halo profiles. 
We notice that even if the $\gamma\gamma$ channel is absent (up to finite terms that we 
neglect here) because of the vanishing mixed anomaly with the $\Uem$, 
as is the case for $\UBpL$ and $\UL$, the $Z\gamma$ channel can be present, because 
it is controlled by the mixed anomaly with the hypercharge.
Most monochromatic photons come from DM annihilation to $\gamma \gamma$ when it is present, 
as the $Z \gamma$ mode is less common and provides half as many photons per annihilation. 
 In the $\Upaxt$ we include the $Z \gamma$ channel above $m_\chi \gtrsim 140$~GeV, where the difference in energies between photons from $\gamma \gamma$ and $Z \gamma$ annihilations is expected to be below the resolution of Fermi-LAT; that of HESS is 
 worse. 
 For simplicity, below this threshold we ignore annihilations to $Z \gamma$, which should not significantly affect our final results due to the lower cross section for this channel.

\begin{figure}[tbp]
\centering
\includegraphics[width=0.8\textwidth]{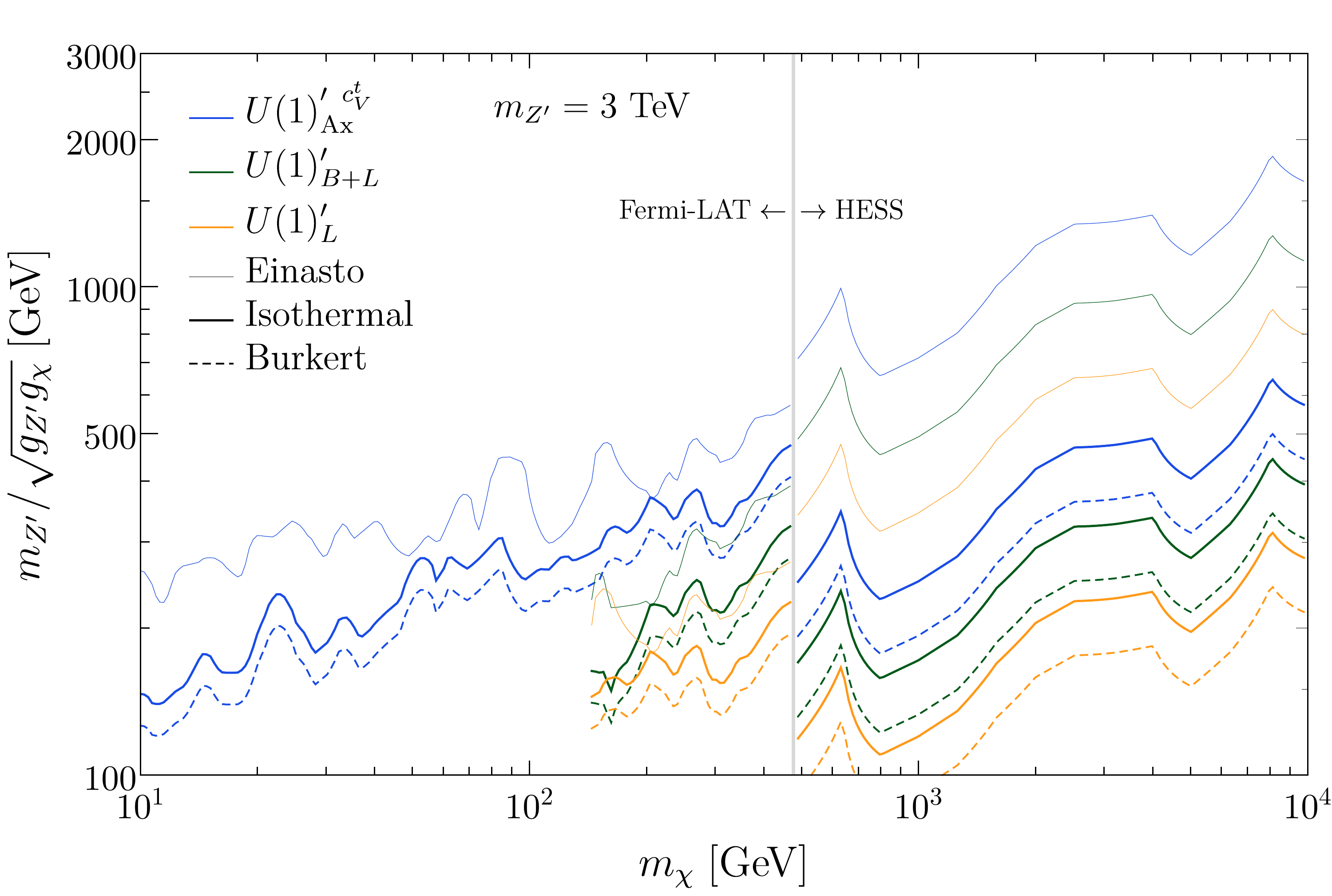}
\caption{\label{fig:linelimits} Limits from Fermi-LAT~\protect{\cite{Ackermann:2015lka}} and HESS~\protect{\cite{Abramowski:2013ax}} searches for $\gamma$ ray lines. Below (above) DM masses of 500 GeV, Fermi-LAT (HESS) provides the constraint. The choice of mediator mass affects only the resonance region $m_\chi \approx m_{Z'} / 2$.}
\end{figure}

The resulting constraints are presented in Fig.~\ref{fig:linelimits}, 
and they clearly illustrate the impact of the anomalies on indirect detection constraints. 
Conversely, anomaly-free models often do not face meaningful limits from 
$\gamma$ ray line searches, due to suppressed annihilation 
cross sections to photons~\cite{Jacques:2016dqz}. 
In the two models $\UBpL$ and $\UL$, where only $Z\gamma$ contributes to the signal, 
the final bound is clearly much weaker, but still non-negligible for a DM mass of a few TeV.
The limits are quite sensitive to the choice of halo profile, particularly for HESS which 
presents limits for a $\gamma$ ray line search in a very narrow region around the Galactic Center. As expected, the best limits are obtained for the cuspy Einasto profile. The sensitivity of line searches is considerably weakened in the resonance region, where annihilation 
to $\gamma \gamma$ is forbidden due to Landau-Yang theorem and only the 
$Z \gamma$ mode contributes.

\subsubsection{Neutrino telescopes}
We finally consider DM annihilation to neutrinos in the Sun, and the associated bounds from  three years of observations of
IceCube~\cite{Aartsen:2016zhm}. In particular, annihilations to $W^+ W^-$, 
$ZZ$ and $\tau^+ \tau^-$ produce high-energy neutrinos which are tightly constrained, while annihilations to $b \bar{b}$ are less strongly limited because of the softer neutrino spectrum. 
To obtain the limit on the overall annihilation cross section -- and hence the scale of DM-SM interactions -- in any given model, 
we must convolve the various IceCube limits on different annihilation channels with the annihilation branching ratios in the model, as we did above for the continuum $\gamma$ ray 
bounds. 

For DM that is captured in the Sun, the typical kinetic energy is of the same order as the temperature at the center of the Sun, $10^7~\mathrm{K}\sim$~keV, which corresponds to 
negligible velocity for DM heavier than the MeV scale. At such energies, $p$-wave annihilation is essentially non-existent, and the annihilation branching ratios are similar to those in the
 Milky Way halo shown previously in Fig.~\ref{fig:brhalo}.\footnote{The only 
 difference between the annihilation branching ratios at velocities characteristic of the Milky Way halo and of the center of the Sun is that the resonance region is narrower in the latter case, due to the even smaller average DM velocity.}

The IceCube bounds on annihilations into $b\bar{b}$ are weaker than the bounds 
on annihilations into $ \tau^+ \tau^-$ by 2-3 orders of magnitude. 
Therefore at low DM mass $\tau^+ \tau^-$ annihilations always provide the most 
constrained source of neutrinos, even in the $\Upaxt$ model when bottom quarks are the
 main products of DM annihilation. 

At higher masses, $W^+ W^-$ and $ZZ$ annihilations face bounds from IceCube that are nearly as strong as $\tau^+ \tau^-$~\cite{Aartsen:2016zhm}, and annihilations to $Z \gamma$ produce half as many neutrinos as $ZZ$. 
Thus we use the stronger of the bounds on annihilations to $\tau^+ \tau^-$ and 
$W^+ W^-, ZZ, Z\gamma$, scaling the IceCube limits by the appropriate branching ratios 
and assuming that the neutrino spectrum for the latter channels are all similar to that for $W^+ W^-$. 

The translation of the IceCube bounds on the SD DM-proton scattering cross section $\sSD$ to bounds on the EFT scale $\mZp/\sqrt{\gZp \gX}$ requires some care about the form factor assumed for DM capture in the Sun. 
When providing a bound on $\sSD$, IceCube assumes that DM and the SM interact through the NR operator $\mathcal{O}_4^\text{NR}$ (according to the standard notation, see e.~g.~\cite{DelNobile:2013sia,Catena:2015uha}).
This is indeed the operator that arises in the $\Upaxt$ model, but for the $\UBpL$ model, the leading interaction is the SI velocity-suppressed operator $\mathcal O_8^\text{NR}$.
To convert between bounds on these operators, we use the capture form factors provided by \cite{Catena:2015uha}.
In the leptophilic model $\UL$ the DM capture rate is negligible, given the small momentum exchange between DM and free electrons in the Sun, and the suppressed loop interaction with nucleons, 
so IceCube bounds do not apply.

\begin{figure}[tbp]
\centering
\includegraphics[width=0.8\textwidth]{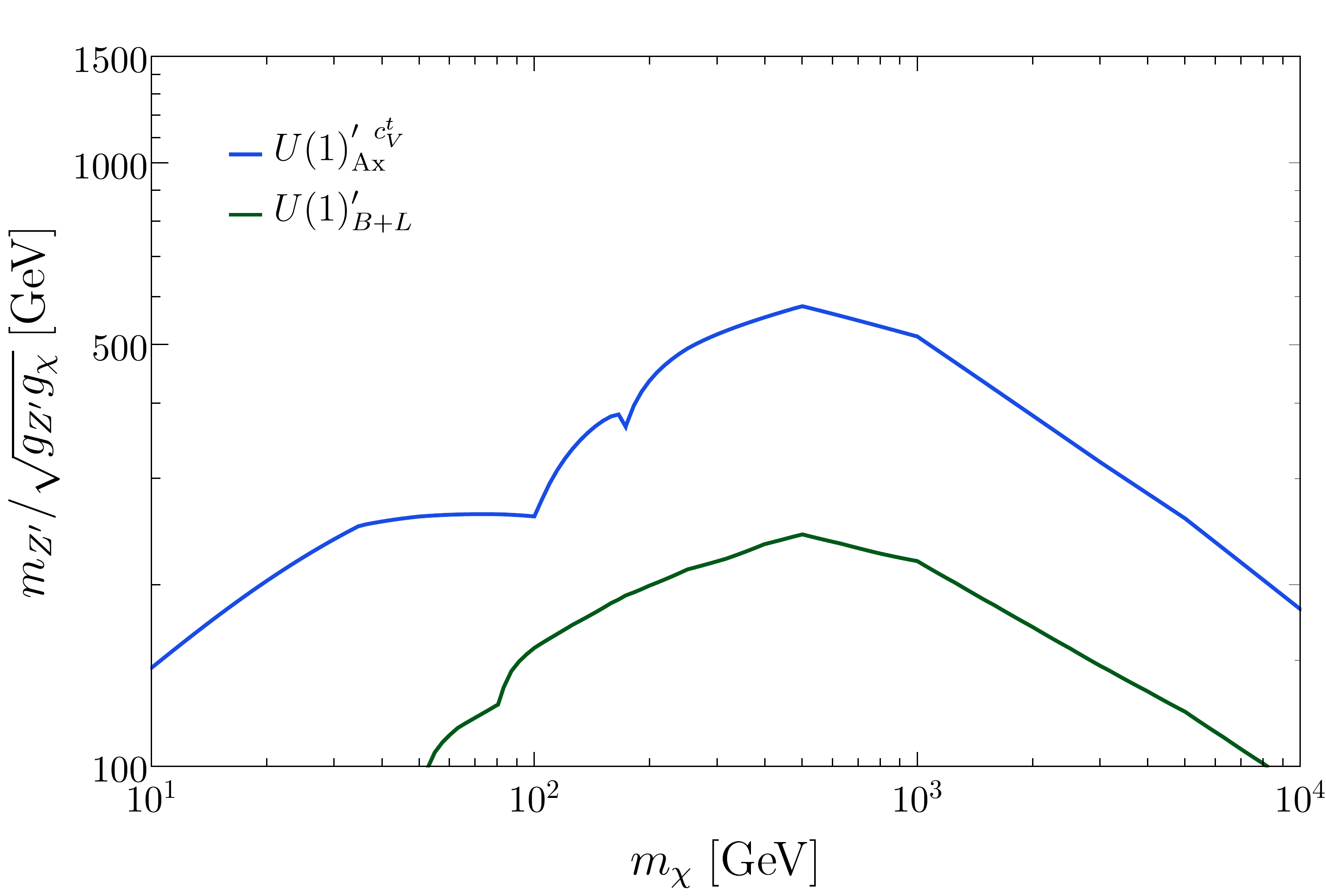}
\caption{\label{fig:icecube} Bounds due to IceCube~\protect{\cite{Aartsen:2016zhm}} 
searches for neutrinos originating from DM annihilations in the Sun, 
on the models we consider.}
\end{figure}

The results are shown in Fig.~\ref{fig:icecube}. 
Because the IceCube bounds are sensitive to the branching ratios of DM annihilations 
rather than to the absolute annihilation cross sections, 
in the $\Upaxt$ model the bounds are weakened due to the large branching ratio into gluons, which yield a negligible neutrino spectrum.
In the $\UBpL$ model, while there are no annihilations into gluons, the velocity-suppressed capture rate results in an even looser bound.
We will see in the next section that in this model, direct detection bounds are much stronger due to coherent enhancement of the 
spin-independent scattering cross section.

\subsection{Colliders and direct detection}

In addition to the above indirect detection searches which can be significantly affected 
by the presence of anomalies, simplified models of DM face complementary constraints from collider and direct detection experiments. In order to present a complete account of the limits on 
the models we consider, here we discuss these bounds and compare them to the exclusions 
derived previously that rely on anomalies.

At the LHC, the main probes of simplified DM models are missing energy-based searches, such as monojets and monophotons, and direct searches for the mediator decaying to SM particles.

\begin{figure}[tbp]
\centering
\includegraphics[width=0.8\textwidth]{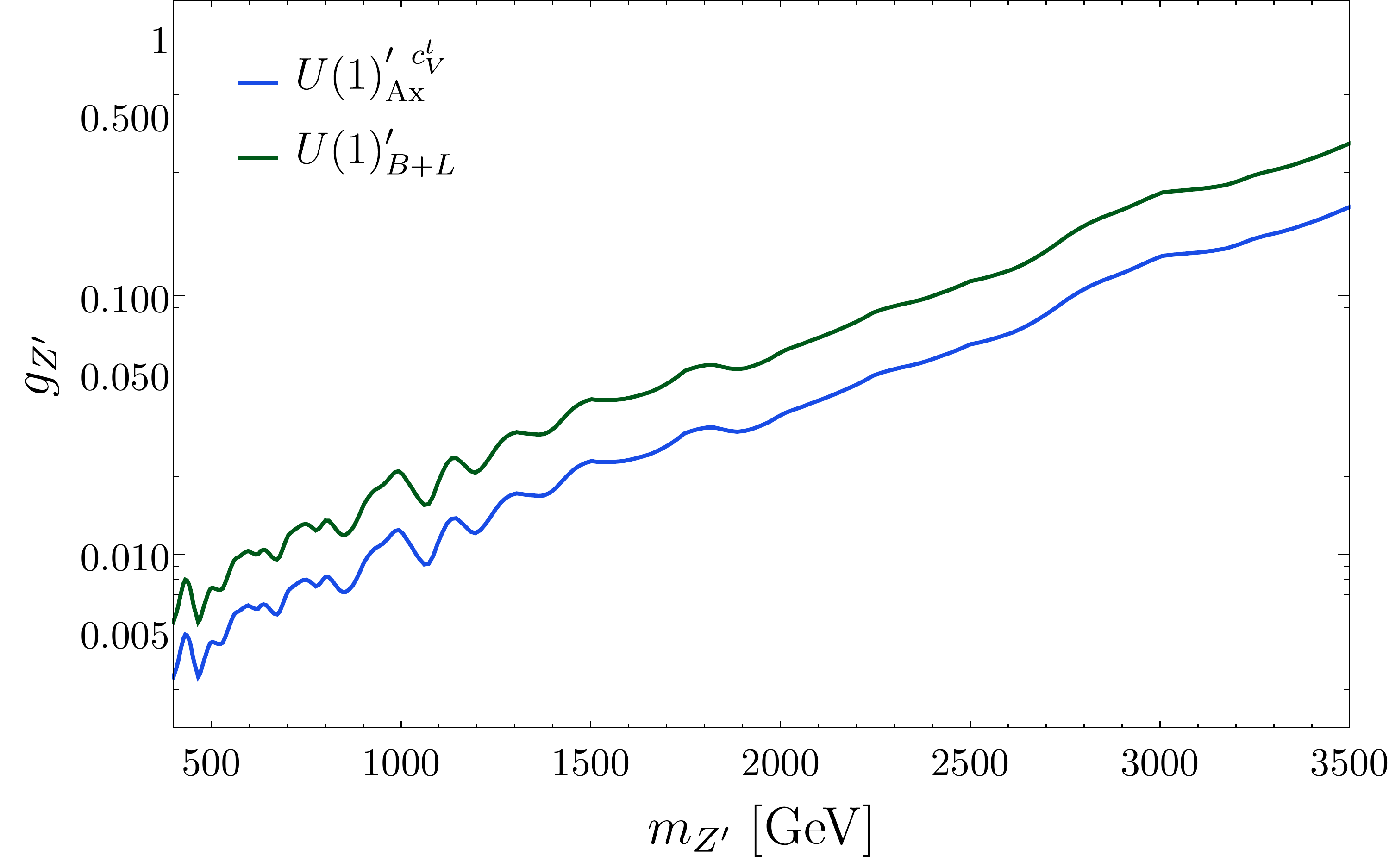}
\caption{\label{fig:lhcdirect} Bounds from resonant dilepton searches on a $Z'$ at LHC 
from the 8~+~13~TeV combined analysis~\protect{\cite{Khachatryan:2016zqb}}, 
presented in the $\gZp$-$\mZp$ plane for $\Upaxt$ and $\UBpL$ 
($\UL$ is obviously unconstrained).}
\end{figure}

The stronger constraints from direct $Z'$ searches come from searches for dilepton resonances. We use the combined 8~+~13~TeV CMS dilepton analysis~\cite{Khachatryan:2016zqb}. 
Because the resonant mediator searches do not involve the DM-mediator coupling, their reach cannot be presented in terms of the DM-SM interaction suppression scale without additional assumptions. Instead, we choose to show in Fig.~\ref{fig:lhcdirect} the upper limit on the $U(1)'$ coupling as a function of the mediator mass. 
The bound on the $\UBpL$ model is rescaled to account for the different charge of light quarks in this model. 
In the leptophilic model $\UL$ the LHC bound does not apply, since the production of $Z'$ 
at the LHC is absent at tree level.
The bound from the dilepton searches for the $Z'$ is quite strong for mediators 
that are kinematically accessible, and would push us to very low $Z'$ couplings to SM fermions. 

\begin{figure}[tbp]
\centering
\includegraphics[width=0.8\textwidth]{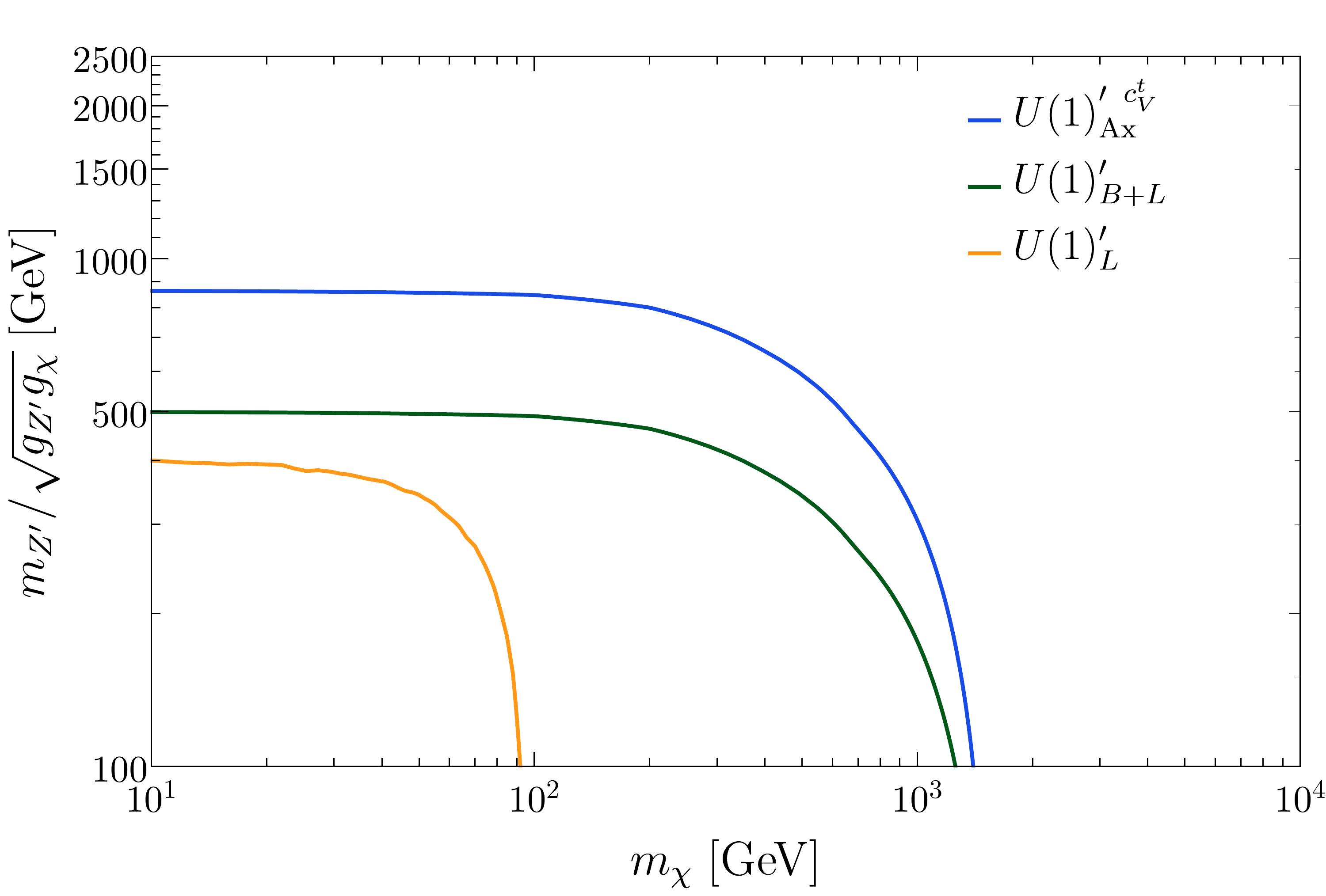}
\caption{\label{fig:monoX} Bounds from the CMS monojet search~\protect{\cite{Khachatryan:2014rra}} 
(for the models $\Upaxt$ and $\UBpL$), and the monophoton search performed by DELPHI and recast 
in~\protect{\cite{Fox:2011fx}} for the $\UL$ model.}
\end{figure}

We also rescale the limits of the 8 TeV CMS monojet 
search~\cite{Khachatryan:2014rra} for Majorana DM and show the results in 
Fig.~\ref{fig:monoX}.\footnote{While more recent searches are available, they are more 
difficult to recast for our purposes. The inclusion of 13~TeV results would improve 
the monojet limits at light DM mass in Fig.~\ref{fig:monoX}, while leaving 
the situation unchanged for DM heavier than several hundred GeV.} For DM below the TeV scale, monojets provide the dominant bound on the models that we consider. 

The LHC monojet analysis clearly does not apply in the $\UL$ model, for which we rely on 
the recast done in~\cite{Fox:2011fx} of the monophoton + missing transverse energy 
searches performed by the DELPHI collaboration at LEP~\cite{Abdallah:2003np,Abdallah:2008aa}. 
Due to the lower energy reach of LEP, the exclusion limit extends up to $\mX\sim 100$ GeV.

Two models, out the three we consider, also produce direct detection signatures. 
The $\Upaxt$ model mainly produces spin-dependent interactions because of the axial couplings\footnote{Spin-independent direct detection is in principle induced at loop level~\cite{Haisch:2013uaa}. However, for typical DM and mediator masses in our region of interest the associated cross section is small enough to be safely ignored.}. 
The most powerful direct detection bound for our purposes comes from PICO~\cite{Amole:2017dex}, and is shown in Fig.~\ref{fig:combination axt}. 
The bound is comparable to the monojet exclusion limit, and is superseded by Fermi-LAT observations of dSph at DM masses around 500~GeV.

The $\UBpL$ model induces instead a spin-independent and velocity-suppressed interaction. 
The most recent experimental bound comes from XENON-1T~\cite{Aprile:2017iyp}. 
The collaboration provides a limit obtained assuming that the interaction between DM and 
nuclei occurs via the canonical spin-independent operator $\mathcal O_1^q$ in the NR EFT of the DM.
In our case the dominant interaction is $\mathcal O_6^q$ rather than 
$\mathcal O_1^q$~\cite{Fitzpatrick:2012ix,Fitzpatrick:2012ib,Anand:2013yka}.
We perform a recast by means of the tables 
provided by~\cite{DelNobile:2013sia}. The result is shown in Fig.~\ref{fig:combination B+L}. 
The exclusion limit from direct detection is the most powerful for $\mX$ up to a few TeV, where it is superseded by $\gamma$ ray line searches only if we assume a cuspy profile of the DM density distribution like Einasto.

\subsection{Summary of results}

\begin{figure}[t]
\centering
\includegraphics[width=.9\textwidth]{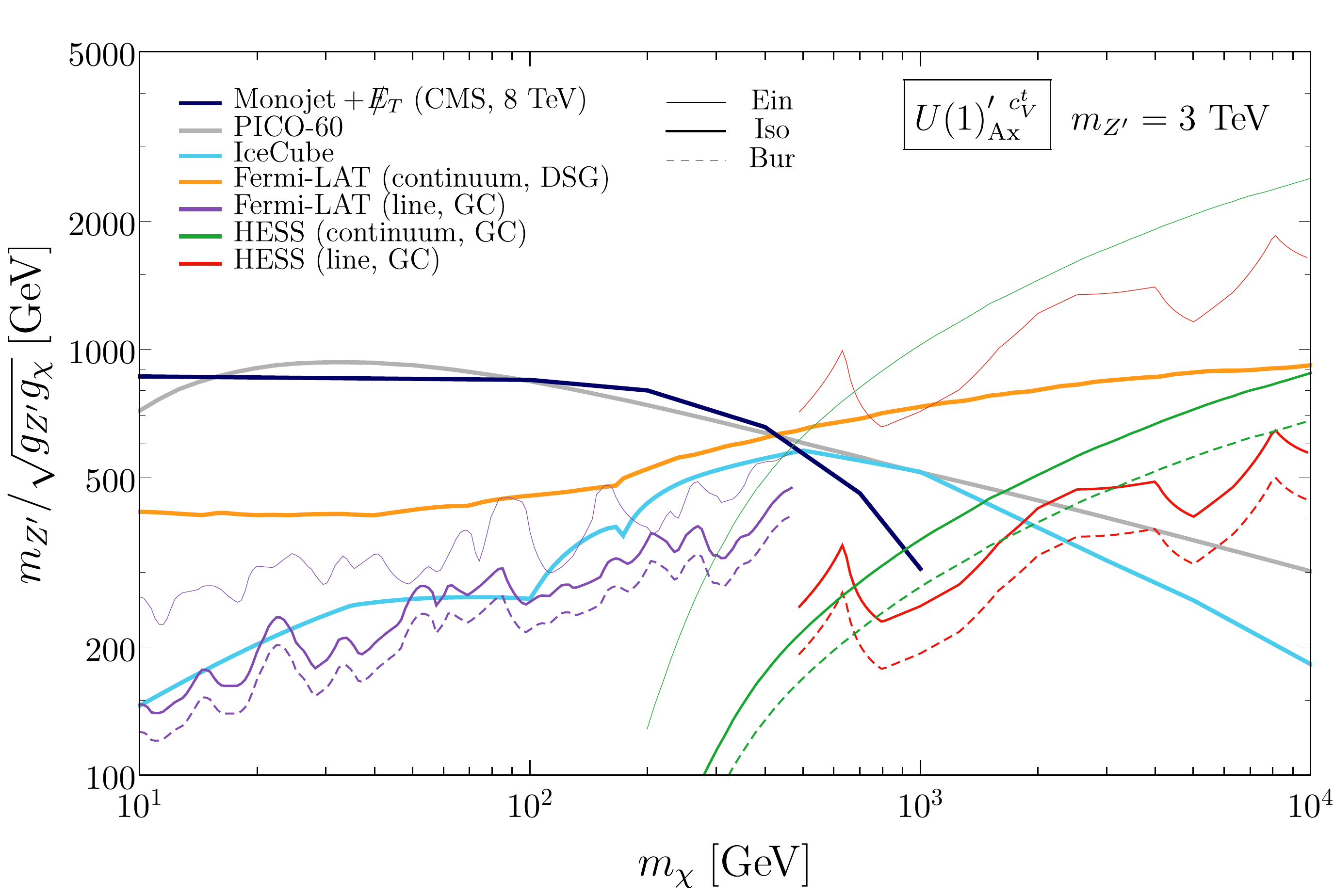}
\caption{\label{fig:combination axt} Combined limits from indirect detection, collider, and direct detection bounds on the $\Upaxt$ model with a 3~TeV mediator. For heavy DM, 
the anomaly-induced annihilations to gauge bosons lead to strong indirect detection bounds. Some of the indirect detection limits are sensitive to the halo profile, and for these the impact of choosing different halo profiles is shown.}
\end{figure}

\begin{figure}[t]
\centering
\includegraphics[width=.9\textwidth]{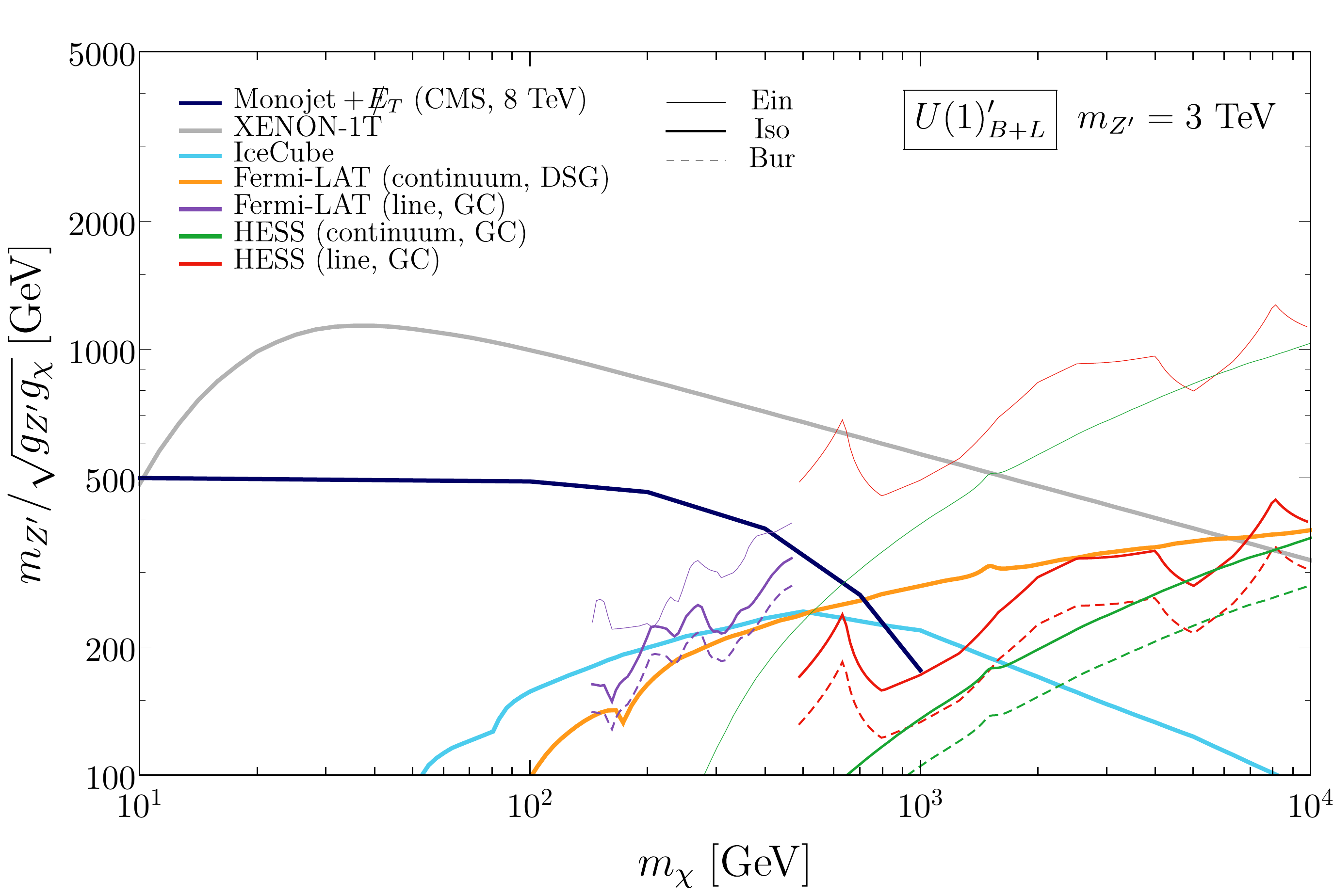}
\caption{\label{fig:combination B+L} The same as in Fig.~\ref{fig:combination axt}, 
shown for the $\UBpL$ model.}
\end{figure}

\begin{figure}[t]
\centering
\includegraphics[width=.9\textwidth]{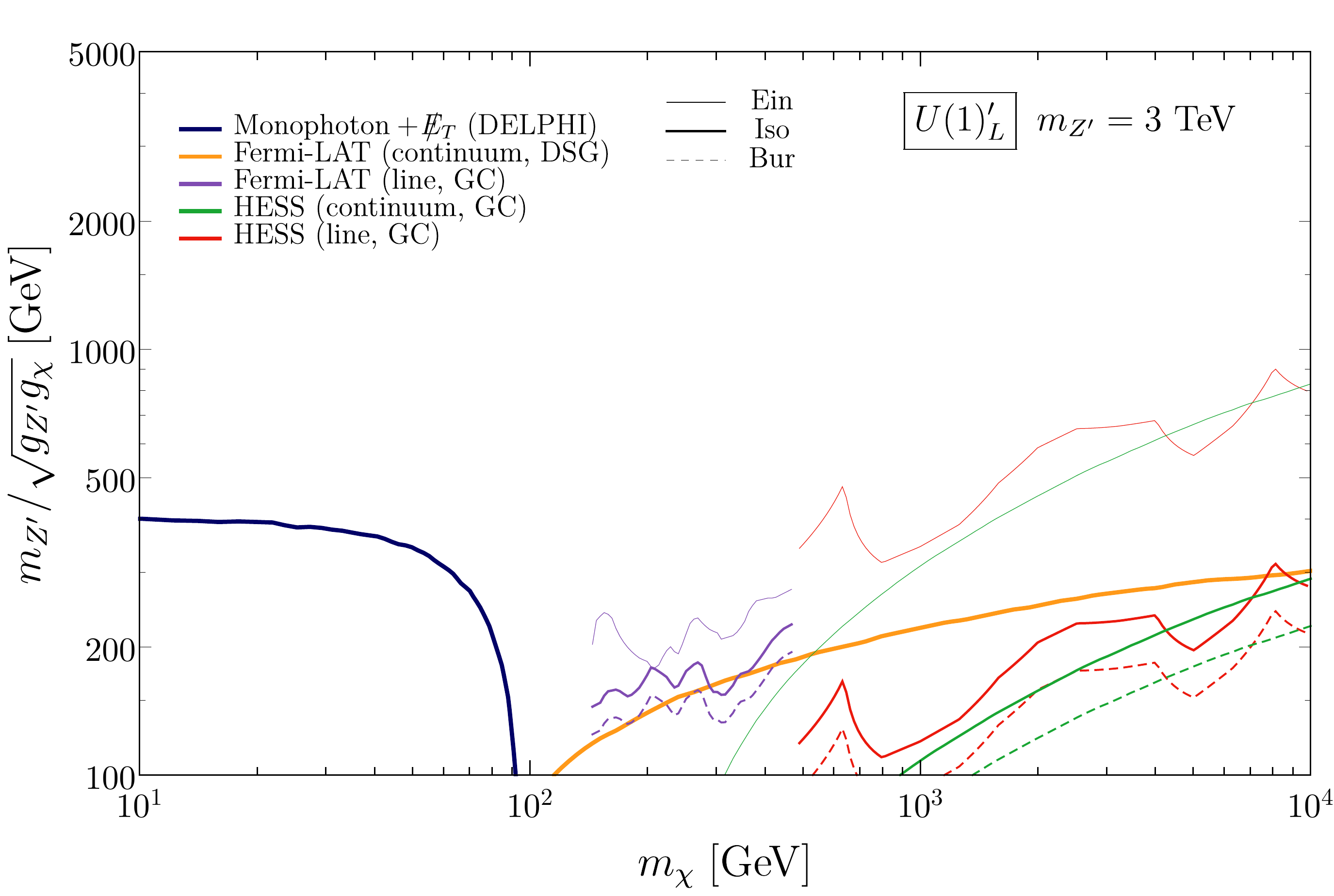}
\caption{\label{fig:combination L} The same as in Fig,~\ref{fig:combination axt}, 
shown for the $\UL$ model.}
\end{figure}

The combinations of all the constraints described above are shown in 
Figs.~\ref{fig:combination axt},~\ref{fig:combination B+L} and~\ref{fig:combination L} 
respectively for the three models $\Upaxt$, $\UBpL$ and $\UL$.

 Indirect detection provides the strongest bounds at large DM mass, driven by loop annihilations of DM to gauge bosons. 
 Depending on the choice of halo profile, either the HESS $\gamma$ ray 
 line search or the Fermi-LAT dSph continuum $\gamma$ ray spectrum analysis is most 
 constraining in this regime, depending in ultimate analysis on whether the $gg$ channel is anomaly-induced or not.
 For models where there is no mixed anomaly between the $Z'$ and
 $\Uem$, the $\gamma$ ray line searches are only weakly constraining.
 For lighter DM, monojets and/or direct searches still provide the 
 tightest bound on the interaction scale.
 Notice that in the leptophilic model $\UL$ these constraints are absent, as is the IceCube bound, and the monophoton 
 searches performed at LEP have a lower reach in $\mX$. In this model, the only limits above LEP are provided by $\gamma$ 
 ray searches.

Outside of the resonance region, the limits are independent of the mediator mass. Consequently, the dilepton searches presented on Fig.~\ref{fig:lhcdirect} 
should be considered as an orthogonal bound to those in Fig.~\ref{fig:combination axt}. For very heavy mediators, on the other hand, only large couplings can currently be constrained. However, future experiments will probe regions of our model which can more naturally accommodate a $Z'$ weighing several TeV, and at such mass scales resonant LHC searches lose sensitivity quite rapidly.

\section{Comments on Validity of Our Results}
\label{sec:validity}

Throughout our discussion, we have assumed that anomalies are canceled 
by fermions that are sufficiently decoupled so as to be effectively infinitely heavy for the 
processes that they mediate. Since the main effects of the Wess-Zumino terms are in 
DM annihilation, this corresponds to $m_f \gg m_\chi$. It is instructive to ask how our results change as the anomaly-canceling fermions are brought closer to the DM mass. 

Let us illustrate this point with the particular example of
DM annihilation in a $U(1)'$ model where the charge of every SM fermion is equal to 
its usual hypercharge. 
Above the scale of the heaviest SM fermion, the top quark, there are no mixed 
$U(1)'$-SM anomalies. Below the top mass, however, anomalies should appear and 
induce Wess-Zumino terms. At some point, where the DM becomes sufficiently lighter than the 
top mass, the EFT should give a good approximation to the full anomaly-free theory.

\begin{figure}[t]
\centering
\includegraphics[width=.8\textwidth]{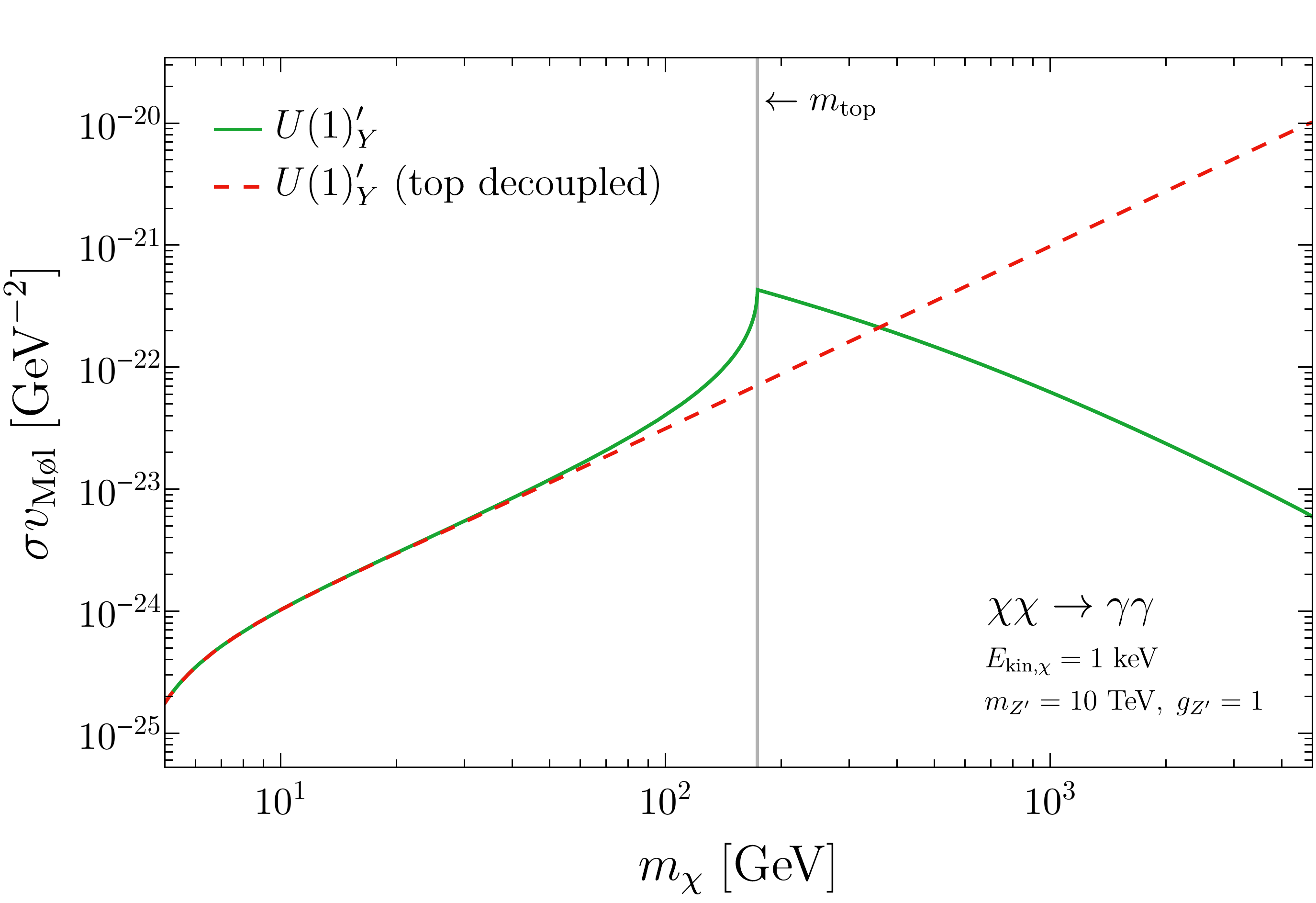}
\caption{\label{fig:anomalon} The effect of decoupling the top quark in a sequential hypercharge model. The green (solid) curve shows the cross section for DM annihilation to photons with $m_t = 175$~GeV, while the red (dashed) curve shows the same cross section with an infinite top quark mass. In general, anomaly-induced effects rise with energy until the mass scale where the anomaly is resolved.}
\end{figure}

We compare these two calculation methods in Fig.~\ref{fig:anomalon}, by 
varying the mass of the DM.
The solid curve in Fig.~\ref{fig:anomalon}  shows the annihilation cross section,
calculated in the full UV complete theory, while the dashed line stands
for the EFT calculation.  
By comparing the two curves, we see that the anomaly-canceling fermions can be 
treated as having infinite mass so long as they are at least 2-3 times 
heavier than the CM energy of the process being studied. In principle it is a very optimistic 
conclusion, that suggests that as long as the spectator fermions are not at the scale of the 
DM, our results are valid.

We also notice that in this particular example we have chosen the mass of the $Z'$ to be 
very high, 10~TeV. Even though the top mass is much smaller than $m_Z'$, the EFT is clearly 
valid in the case of light DM, showing again that the scale of the $Z'$ plays no
role in setting the validity range of the EFT.

\section{Conclusions}
\label{sec:conc}

Simplified models of DM are frequently used to present experimental results, yet the most common spin-1 mediator models often contain anomalies. While these may be resolved at high scales through the introduction of additional chiral fermions, in this work we have demonstrated that this is not without consequence. Integrating out heavy fermions generates Wess-Zumino terms, whose derivative couplings can create significant effects at high energies despite the loop suppression.

In particular, mixed anomalies cause couplings between the $Z'$ and the SM gauge bosons. These interactions affect DM annihilation through the $Z'$, and have the most impact on indirect detection probes of DM. We have evaluated the resulting bounds for a selection of $U(1)'$ possibilities. The Wess-Zumino terms that we have computed depend only on the anomaly coefficients, and so DM annihilation cross sections to gauge bosons for an arbitrary $U(1)'$ can in principle be obtained by scaling our results. If a new $U(1)$' has vector couplings, the only anomaly-induced terms involve the $SU(2)_L$ bosons, and so the annihilation cross sections tend to be smaller. This leads to weaker constraints from indirect detection searches involving photons, under the assumption that the DM relic density is set by some external mechanism.

We have compared bounds from indirect detection with those from direct detection and colliders. We find that $\gamma$-ray searches can often provide the most stringent limits on heavy DM, with either continuum or line searches being more constraining depending on the choice of halo profile. For intermediate masses between a few hundred GeV and 1 TeV, IceCube can provide bounds comparable to direct searches if the scattering cross section with protons is SD. At small DM mass, direct detection is more effective at limiting a $Z'$ which couples to quarks. Monojet and monophoton bounds can also constrain lighter DM, and while resonance searches are not directly comparable, dileptons still provide the best bounds if the mediator is kinematically accessible at the LHC and couples to quarks.

While most of our calculations assumed that the fermions which cancel anomalies are completely decoupled, we also considered the effect of restoring gauge invariance at smaller scales. As long as the anomalies persist up to energies that are a few times higher than the DM mass, which is the relevant energy scale for annihilation, our results remain completely valid.

In our study we have assumed that DM is a Majorana fermion, in part to emphasize our new indirect detection limits over the usual direct detection bounds, which are strong for spin-independent interactions that arise when the DM and quarks both couple vectorially to the $Z'$. It would nevertheless be interesting to examine the interplay between direct and indirect detection bounds in more general models. For instance, if the DM is a Dirac fermion with a vector $U(1)'$ coupling but the SM quarks couple axially under $U(1)'$, the leading spin-independent direct detection interaction is velocity-suppressed. Dressing such an interaction with Higgses yields a pure vector interaction, but as the main effect involves a top loop, it can be avoided if the top does not couple to the $Z'$. On the other hand, if the top does carry $U(1)'$ charge but the light quarks do not, the $SU(3)_C^2 \times U(1)'$ anomaly could be relevant for collider searches as there is no tree-level DM production from light quark initial states.

In characterizing the sensitivities of DM searches, models that are employed to show experimental results should be consistent with theoretical considerations. In addition to the recently well-studied requirement that such models provide unitary scattering amplitudes, we have shown here how gauge invariance necessitates the inclusion of additional interactions beyond the minimal Lagrangian of generic simplified DM models. We look forward to future developments in this direction as searches for DM continue.

\appendix
\section{Effective triple gauge boson couplings}
\label{sec:results}

Equation~\ref{eq:rosenberg} gives the effective $Z'$-$\gamma$-$\gamma$ vertex. Here, we provide the form of this vertex for other gauge boson channels.

The calculation of the $Z'$-$g$-$g$ vertex is the same as for $Z'$-$\gamma$-$\gamma$ up to a color factor and coupling constants:
\beq
\Gamma_{gg}^{\mu\nu\rho} = 2 \left( \frac{g_s}{e Q_f^{em} N_c} \right)^2 \Gamma^{\mu\nu\rho}
\eeq

For massive gauge bosons, we include the Goldstone amplitude in the Ward identities, as described in Sec.~\ref{sec:eftanomaly}. Unlike the photon and gluon cases, a triangle vertex arises even if the $U(1)'$ coupling of the loop fermion is vector-like, because the weak interactions violate parity. Similarly to the $Z'$ coupling to fermions, we write the $Z$-fermion-fermion vertex as $i \frac{g_w}{c_w} \gamma^\rho (g_V^Z + g_A^Z \gamma^5)$. Then, the $Z'$-$Z$-$\gamma$ vertex is given by
\beqa
\label{eq:rosenbergzgamma}
\Gamma^{\mu\nu\rho}_{Z\gamma} &=& \frac{\gZp N_c^2 g_w e Q_f^{em} (g_V g_A^Z + g_A g_V^Z)}{\pi^2 c_w} \Big( I_1^{Z\gamma} \epsilon^{\alpha\nu\rho\mu} (p_2)_\alpha + I_2^{Z\gamma} \epsilon^{\alpha\nu\rho\mu} (p_3)_\alpha \nonumber \\
&&\ + I_3^{Z\gamma} \epsilon^{\alpha\beta\nu\mu} (p_2)^\rho (p_2)_\alpha (p_3)_\beta + I_4^{Z\gamma} \epsilon^{\alpha\beta\nu\mu} (p_3)^\rho (p_2)_\alpha (p_3)_\beta \\
&&\ + I_5^{Z\gamma} \epsilon^{\alpha\beta\rho\mu} (p_2)^\sigma (p_2)_\alpha (p_3)_\beta + I_6^{Z\gamma} \epsilon^{\alpha\beta\rho\mu} (p_3)^\sigma (p_2)_\alpha (p_3)_\beta \Big) \nonumber
\eeqa
where the form factors, in terms of those in Eq.~\ref{eq:formfactors}, are
\begin{small}
\beqa
I_1^{Z\gamma}(p_2, p_3; m_f) &=& (p_2 \cdot p_3) I_3^{Z\gamma}(p_2, p_3; m_f) + p_3^2 I_4^{Z\gamma}(p_2, p_3; m_f) \nonumber \\
I_2^{Z\gamma}(p_2, p_3; m_f) &=& p_2^2 I_5^{Z\gamma}(p_2, p_3; m_f) + (p_2 \cdot p_3) I_6^{Z\gamma}(p_2, p_3; m_f) \nonumber \\
&&\ - \frac{g_V g_A^Z}{g_V g_A^Z + g_A g_V^Z} m_f^2 C_0(p_3^2, p_1^2, p_2^2, m_f^2, m_f^2, m_f^2) \nonumber \\
I_3^{Z\gamma}(p_2, p_3; m_f) &=& -I_3(p_2, p_3; m_f) \\
I_4^{Z\gamma}(p_2, p_3; m_f) &=& -I_4(p_2, p_3; m_f) \nonumber \\
I_5^{Z\gamma}(p_2, p_3; m_f) &=& -I_5(p_2, p_3; m_f) \nonumber \\
I_6^{Z\gamma}(p_2, p_3; m_f) &=& -I_6(p_2, p_3; m_f) \nonumber
\eeqa
\end{small}

The $Z'$-$Z$-$Z$ vertex is
\beqa
\label{eq:rosenbergzz}
\Gamma^{\mu\nu\rho}_{ZZ} &=& \frac{\gZp N_c^2 g_w^2 (2 g_V g_V^Z g_A^Z + g_A ((g_V^Z)^2 + (g_A^Z)^2))}{\pi^2 c_w^2} \Big( I_1^{ZZ} \epsilon^{\alpha\nu\rho\mu} (p_2)_\alpha + I_2^{ZZ} \epsilon^{\alpha\nu\rho\mu} (p_3)_\alpha \nonumber \\
&&\ + I_3^{ZZ} \epsilon^{\alpha\beta\nu\mu} (p_2)^\rho (p_2)_\alpha (p_3)_\beta + I_4^{ZZ} \epsilon^{\alpha\beta\nu\mu} (p_3)^\rho (p_2)_\alpha (p_3)_\beta \\
&&\ + I_5^{ZZ} \epsilon^{\alpha\beta\rho\mu} (p_2)^\sigma (p_2)_\alpha (p_3)_\beta + I_6^{ZZ} \epsilon^{\alpha\beta\rho\mu} (p_3)^\sigma (p_2)_\alpha (p_3)_\beta \Big) \nonumber
\eeqa
where the form factors are
\begin{small}
\beqa
I_1^{ZZ}(p_2, p_3; m_f) &=& (p_2 \cdot p_3) I_3^{ZZ}(p_2, p_3; m_f) + p_3^2 I_4^{ZZ}(p_2, p_3; m_f) - m_f^2 C_0(p_3^2, p_1^2, p_2^2, m_f^2, m_f^2, m_f^2) \nonumber \\
I_2^{ZZ}(p_2, p_3; m_f) &=& p_2^2 I_5^{ZZ}(p_2, p_3; m_f) + (p_2 \cdot p_3) I_6^{ZZ}(p_2, p_3; m_f) - m_f^2 C_0(p_3^2, p_1^2, p_2^2, m_f^2, m_f^2, m_f^2) \nonumber \\
I_3^{ZZ}(p_2, p_3; m_f) &=& -I_3(p_2, p_3; m_f) \\
I_4^{ZZ}(p_2, p_3; m_f) &=& -I_4(p_2, p_3; m_f) \nonumber \\
I_5^{ZZ}(p_2, p_3; m_f) &=& -I_5(p_2, p_3; m_f) \nonumber \\
I_6^{ZZ}(p_2, p_3; m_f) &=& -I_6(p_2, p_3; m_f) \nonumber
\eeqa
\end{small}

For the $Z'$-$W^+$-$W^-$ vertex, we assume that two fermions run in the loop whose left-handed components are related by $SU(2)_L$, with the up-type fermion having mass $m_f$ and coupling vectorially to the $Z'$, $-i \gZp \gamma^\nu$. Then, regardless of whether the down-type fermion coupling to the $Z'$ is vector or axial, i.e. $-i \gZp \gamma^\nu$ or $i \gZp \gamma^\nu \gamma^5$, the vertex is given by
\beqa
\label{eq:rosenbergww}
\Gamma^{\mu\nu\rho}_{WW} &=& \frac{\gZp N_c^2 g_w^2}{4 \pi^2} \Big( I_1^{WW} \epsilon^{\alpha\nu\rho\mu} (p_2)_\alpha + I_2^{WW} \epsilon^{\alpha\nu\rho\mu} (p_3)_\alpha \nonumber \\
&&\ + I_3^{WW} \epsilon^{\alpha\beta\nu\mu} (p_2)^\rho (p_2)_\alpha (p_3)_\beta + I_4^{WW} \epsilon^{\alpha\beta\nu\mu} (p_3)^\rho (p_2)_\alpha (p_3)_\beta \\
&&\ + I_5^{WW} \epsilon^{\alpha\beta\rho\mu} (p_2)^\sigma (p_2)_\alpha (p_3)_\beta + I_6^{WW} \epsilon^{\alpha\beta\rho\mu} (p_3)^\sigma (p_2)_\alpha (p_3)_\beta \Big) \nonumber
\eeqa
where the form factors are
\begin{small}
\beqa
I_1^{WW}(p_2, p_3; m_f) &=& (p_2 \cdot p_3) I_3^{WW}(p_2, p_3; m_f) + p_3^2 I_4^{WW}(p_2, p_3; m_f) - \frac{m_f^2}{4} \Big( C_0(p_3^2, p_1^2, p_2^2, m_f^2, 0, 0) \nonumber \\
&&\ + C_1(p_3^2, p_1^2, p_2^2, m_f^2, 0, 0) + C_1(p_3^2, p_1^2, p_2^2, 0, m_f^2, m_f^2) \nonumber \\
&&\ + C_2(p_3^2, p_1^2, p_2^2, m_f^2, 0, 0) + C_2(p_3^2, p_1^2, p_2^2, 0, m_f^2, m_f^2) \Big) \nonumber \\
I_2^{WW}(p_2, p_3; m_f) &=& p_2^2 I_5^{WW}(p_2, p_3; m_f) + (p_2 \cdot p_3) I_6^{WW}(p_2, p_3; m_f) + \frac{m_f^2}{4} \Big( C_0(p_3^2, p_1^2, p_2^2, m_f^2, 0, 0) \nonumber \\
&&\ + C_1(p_3^2, p_1^2, p_2^2, m_f^2, 0, 0) + C_1(p_3^2, p_1^2, p_2^2, 0, m_f^2, m_f^2) \nonumber \\
&&\ + C_2(p_3^2, p_1^2, p_2^2, m_f^2, 0, 0) + C_2(p_3^2, p_1^2, p_2^2, 0, m_f^2, m_f^2) \Big) \nonumber \\
I_3^{WW}(p_2, p_3; m_f) &=& \frac{1}{2} \left( C_{12}(p_2^2, p_1^2, p_3^2, m_f^2, 0, 0) + C_{12}(p_2^2, p_1^2, p_3^2, 0, m_f^2, m_f^2) \right) \\
I_4^{WW}(p_2, p_3; m_f) &=& -I_5^{WW}(p_3, p_2; m_f) \nonumber \\
I_5^{WW}(p_2, p_3; m_f) &=& \frac{1}{2} \Big( C_{11}(p_2^2, p_1^2, p_3^2, m_f^2, 0, 0) + C_{11}(p_2^2, p_1^2, p_3^2, 0, m_f^2, m_f^2) \nonumber \\
&&\ + C_1(p_2^2, p_1^2, p_3^2, m_f^2, 0, 0) + C_1(p_2^2, p_1^2, p_3^2, 0, m_f^2, m_f^2)  \Big) \nonumber \\
I_6^{WW}(p_2, p_3; m_f) &=& -I_3^{WW}(p_2, p_3; m_f) \nonumber
\eeqa
\end{small}

\acknowledgments

We are grateful to Uli Haisch, Ian Low, Toni Riotto, Andrea Wulzer, Giulia Zanderighi and Jure Zupan for useful discussions, 
and to Mohamed Rameez for useful information about the IceCube analysis.
The work of AI is supported in part by the U.S. Department of Energy under grant No.~DE-SC0015634, and in part by PITT PACC.
DR is supported by the Swiss National Science Foundation (SNSF), project ``Investigating the Nature of Dark Matter'' (project number: 200020\_159223).
AI and AK are grateful to the Aspen Center of Physics,
 which is supported by  grant NSF~PHY-1066293,  where this project was originally 
initiated.
AK is also grateful to the Mainz Institute for Theoretical Physics (MITP) for its hospitality and partial support when this work has been finalized. 

\paragraph{Note added.} While this paper was being completed, related work~\cite{Dror:2017ehi} appeared which concentrates on the implications of anomalous $U(1)'$ theories at low energies. The underlying physics is similar, 
but in contrast to the authors of~\cite{Dror:2017ehi}, we primarily consider theories of WIMP dark matter.

\bibliographystyle{JHEP}
\bibliography{dmanomaly}

\end{document}